\documentclass[twocolumn,amsmath,showpacs,amsfonts,aps,prc,floatfix]{revtex4}
\usepackage{graphicx}
\usepackage{bm}
\usepackage{epsfig}
\begin{document}
\title{2+1 dimensional hydrodynamics including bulk viscosity: a systematic study}            
\author{Victor Roy and A.K.Chaudhuri}
\affiliation{Variable Energy Cyclotron Centre,Kolkata-700064}
\begin{abstract} 

We have studied the effect of nonzero bulk viscosity with peak near the lattice QCD 
predicted crossover temperature $T_{co}\sim 175 MeV$ on   charged particle transverse momentum spectra
and elliptic flow. The Israel-Stewart theory of 2nd order causal dissipative relativistic fluid
dynamics is used to simulate the space time evolution of the matter formed in Au-Au collisions
at $\sqrt{s_{NN}}$=200 GeV assuming longitudinal 
boost invariance. A systematic comparison of temperature, transverse
velocity, spatial and momentum anisotropy evolution of the ideal, bulk and shear 
viscous fluid has been carried out. Two different temperature dependent forms of  $\zeta/s$ and
a constant $\eta/s$ was used. Both the bulk and shear viscous correction to the ideal freezeout distribution 
function are included. The dissipative correction to the freezeout distribution for bulk viscosity
was calculated using Grad's fourteen moment method. From our simulation we show that 
the method is applicable only for $\zeta/s < 0.005\times\eta/s|_{KSS}$ for freezeout temperatures 130 and 160 MeV. 
\end{abstract}
\maketitle

\section{Introduction}

One of the most interesting results coming from RHIC heavy ion program is 
the observation that hot QCD matter created in Au-Au collisions behaves like 
an almost ideal fluid  \cite{Adcox:2004mh,Adams:2005dq,Arsene:2004fa,Gyulassy:2004zy}.

Relativistic hydrodynamics has been a successful theory to describe the 
bulk properties of the QCD medium produced in nuclear collision.
Hydrodynamic simulation of nuclear collision at RHIC indicates
that the shear viscosity of QCD plasma is very small. However the extracted
value of shear viscosity largely depend on the initial condition 
\cite{Hirano:2005xf}.
The theoretical estimation of various kinetic coefficient for a 
QCD plasma becomes enormously complex due to strong coupling.
String theoretical calculation based on the ADS/CFT correspondence
predicts a lower limit on the ratio of shear viscosity to entropy
density as $\eta/s\geq 1/4\pi$ \cite{Kovtun:2004de}.

The bulk viscosity $\zeta$ can in general be of the same order of
magnitude as shear viscosity $\eta$. But until recently, the 
bulk viscosity was neglected in the study of fluid dynamics of
QCD matter. There are important
special cases in which $\zeta$ is very much smaller, or vanishes
altogether \cite{Weinberg:1971mx}. In \cite{Weinberg:1971mx} author
shows that a simple gas of structureless point particles will have
negligible bulk viscosity in the extreme-relativistic or 
non relativistic limits. However, it must be stressed that 
a vanishing bulk viscosity is the exception, rather than 
the rule, for general imperfect fluids \cite{Weinberg:1971mx},\cite{Landau:fluid}.

Existence of non zero bulk viscosity in QCD plasma 
near the critical temperature was reported in \cite{Meyer:2007dy,arXiv:0710.3625,
arXiv:0711.0914} from
lattice QCD calculation. The theoretical estimation for 
transport coefficients in QCD plasma can also be found in
\cite{Danielewicz:1984ww,Kapusta:2008vb,arXiv:0906.5592,nucl-th/0604008,
hep-ph/9409250,362262}.
In  \cite{Danielewicz:1984ww} a lower and upper bound of
shear viscosity was given and bulk viscosity associated with
the plasma to hadron transition was estimated in the relaxation-
time approximation. Several sources of shear and bulk viscosity 
was discussed in~\cite{nucl-th/0604008} with the emphasis that
the bulk viscosity is associated with the chemical nonequilibrium.
The effect of bulk viscosity on freezeout
and HBT puzzle was studied in \cite{Torrieri:2007fb}.
In spite of all these theoretical calculations, there are several
model dependency in the estimate of the transport coefficients. 
One can introduce the dissipative 
effects in hydrodynamics by treating transport coefficient like
shear and bulk viscosity as input parameter and obtain their
values in phenomenological study by comparing to the experimental data.
In the following, we will consider effect of bulk viscosity on hydrodynamic evolution and subsequent particle production.
The temperature dependence of $\zeta/s$ is different 
for leading order pQCD and AdS/CFT calculation \cite{Meyer:2007dy}.
In pQCD calculation $\zeta=15\eta(T)\left(1/3-c^{2}_{s}(T)\right)^{2}$
and in AdS/CFT $\zeta \sim \eta(T)\left(1/3-c^{2}_{s}(T)\right)$, where
$c_{s}$ is the speed of sound in the medium, $\eta$ is the shear
viscosity and T is the temperature. 
$\zeta/s$(T) is different among the available lattice calculations
\cite{arXiv:0711.0914,Meyer:2007dy,arXiv:0710.3625}.
In this study we will describe in detail 
the results obtained from the recently extended version of our code by considering 
two different temperature dependent forms of $\zeta/s$ and a constant $\eta/s$. 
In addition to shear viscous correction to freezeout distribution function we have 
also considered the bulk viscous correction to the ideal freezeout distribution function.
The bulk viscous correction  $\delta f_{bulk}$ was calculated by using Grad's 14-moment method for 
a multicomponent gas as discussed in \cite{Monnai:2009ad}. 
Considering a small value of bulk stress $\Pi$ at the freezeout it was shown in \cite{Monnai:2009ad} 
that bulk viscosity  has a non-negligible effect on particle spectra and
elliptic flow coefficient. Thus it is important to consider the bulk viscous evolution
and the dissipative correction to the ideal freezeout distribution 
function. We will concentrate in this work to study the effect of bulk viscosity on 
$p_{T}$ spectra and elliptic flow of charged hadron with
for two different freezeout temperature $T_{f}$=130 and 160 MeV. 
We want to remind that the calculation of $\delta f_{bulk}$ for a multicomponent hadron gas
is non trivial, for a recent calculation of the dissipative correction $\delta f$ to 
the single particle distribution function in leading-log QCD and in several simplified
model see \cite{arXiv:1109.5181}.

The paper is organized as follows, in section \ref{sec2}, we discuss the formalism
of bulk viscous hydrodynamics in the context of Israel-Stewart 2nd order theory.
Section III deals with the input ($\eta/s,\zeta/s$(T), initial condition, relaxation time,
equation of state and freezeout condition) required for the 2+1D viscous hydrodynamics
simulation. In section IV we present the results of our simulation which includes 
the time evolution of bulk viscous stress $\Pi\left(x,y\right)$ as well as its spatial
average $\left\langle \Pi\left(x,y\right)\right\rangle$, and the effect of $\zeta/s$ on
evolution and observables. A comparative study between ideal, shear and bulk viscous
fluid evolution has also been done. 
The effect of the bulk viscous correction to the ideal 
freeze-out distribution function on charged pion's elliptic flow and $p_{T}$ spectra 
is discussed. Finally in section V we present a summary of our study.

\section{Formulation}\label{sec2}

The most widely studied theory of relativistic causal dissipative hydrodynamics 
is due to Israel-Stewart(I-S)\cite{Israel:1976,nucl-th/0611090,Chaudhuri:2008sj}. 
In the present work we follow the I-S formalism of viscous 
hydrodynamics in 2+1D to study the effect of bulk viscosity on experimental observables.
For completeness we start with the brief description of the formalism which is followed by
a detail discussion on implementing bulk viscosity in the fluid evolution as well as the
corresponding dissipative correction to the freezeout distribution function.

For a simple fluid in equilibrium or in adiabatic limit,
the energy momentum tensor $T^{\mu\nu}$ and particle current $N^{\mu}$ has the 
following forms \cite{Weinberg:1971mx}

\begin{eqnarray}
T^{\mu\nu}_{eq}&=&(\epsilon+p)u^{\mu}u^{\nu}-pg^{\mu\nu}\\
	N^{\mu}_{eq}&=&nu^{\mu}
\end{eqnarray}

where $\epsilon$ ,$p$ and $n$ are  total energy density, pressure, and particle 
number density. $u^{\mu}$ is the velocity four-vector, normalized so that
$u^{\mu}u_{\mu}=1$. Choice of hydrodynamic velocity is arbitrary. There are two commonly used comoving frame to fix the $u$, (i) Landau-Lifshitz
frame where $u^{\mu}$ is parallel to the
energy four flow and (ii)Eckart frame where the $u^{\mu}$ is parallel to
the $N^{\mu}$. We are using the Landau-Lifshitz frame, which is more appropriate than the Eckart frame, in baryon free central rapidity region.

Due to the presence of dissipative processes the energy momentum 
tensor and the particle four flow of ideal fluid gets modified as
follows:

\begin{eqnarray}
		T^{\mu\nu}&=&T^{\mu\nu}_{eq}+\Delta T^{\mu\nu} \nonumber \\
 		N^{\mu}&=&	N^{\mu}_{eq}+\Delta N^{\mu}	         
\end{eqnarray}

$\Delta T^{\mu\nu}$ and $\Delta N^{\mu}$ are dissipative corrections to the energy momentum tensor and particle 4-current.
In the Landau-Lifshitz frame, the dissipative correction to the particle current and energy-momentum tensor can be written as,

\begin{eqnarray} \label{eq5}
\Delta N^{\mu\nu}&=&-n\frac{q^\mu}{\epsilon+p}\\
\Delta T^{\mu\nu}&=&\Pi (u^{\mu}u^{\nu}-g^{\mu\nu})+\pi^{\mu\nu}.
\end{eqnarray}
 
In Eq.\ref{eq5}, $q^\mu$ is the heat conduction current, $\Pi$ is the bulk viscous stress and $\pi^{\mu\nu}$ is the shear stress tensor.
The space time evolution of the fluid is governed by the
  energy momentum and particle number 
conservation laws, 

\begin{eqnarray}
\partial_{\mu}T^{\mu\nu} &=&0,\\
\partial_{\mu}N^{\mu} &=&0.
\end{eqnarray}

Apart from these two conservation laws, the second law of thermodynamics 
restrict the entropy four vector $S^{\mu}$ in such a way that the
rate of entropy production per unit volume should always be positive
for all possible fluid configurations,

\begin{equation}
		\partial_{\mu}S^{\mu} \geq 0
\end{equation}

For an imperfect fluid the entropy four vector $S^{\mu}$
can be decomposed into two parts.

\begin{equation}
	S^{\mu}=S^{\mu}_{eq}+\Delta S^{\mu}
\end{equation}
where $S^{\mu}_{eq}$ is the equilibrium part and 
$\Delta S^{\mu}$ is the dissipative correction. The equilibrium entropy four current is given by,

\begin{equation}
	S^{\mu}_{eq}=\frac{pu^{\mu}-\mu N^{\mu}_{eq}+u_{\nu}T^{\nu\mu}_{eq}}{T}
\end{equation}

If we neglect the heat conduction and consider bulk and shear viscosity as the only dissipation mechanism, the   non-equilibrium correction to the entropy four current can be written as,
 
\begin{eqnarray}
\Delta S^{\mu}= -\frac{\beta_{0}}{2T}u^{\mu}\Pi^{2}-\frac{\beta_{2}}{2T}u^{\mu}\pi_{\alpha\beta}\pi^{\alpha\beta}
\end{eqnarray}

\noindent 
where the co-efficients $\beta_{0}$ and $\beta_{2}$ are function of energy density and
number density. Their exact values can be determined from kinetic theory \cite{Israel:1976}. 
Later we will identify that  these 
co-efficients are related to the relaxation time of bulk and shear viscosity
respectively.

The $\Delta S^{\mu}$ is constructed 
in terms of various orders of the gradient of $u^{\mu}$.
For a first order theory the second law of thermodynamics 
$\partial_{\mu}S^{\mu} \geq 0$ can be satisfied by postulating  
the following form of bulk $\Pi$ and shear $\pi^{\mu\nu}$ stress

\begin{eqnarray}
	\Pi&=&-\zeta\theta \label{eq:NS}\\
 	\pi^{\mu\nu}&=&2\eta\nabla^{<^{\mu}u^{\nu}>}
\end{eqnarray}

\noindent
where $\zeta$ and $\eta$ are the positive transport coefficients,
bulk and shear viscosity respectively. $\theta$ is the expansion
rate to be defined later.

It can be shown that the first order theory violates causality~\cite{Hiscock:1985zz}. 
For example if, in a given fluid cell, thermodynamic forces vanish, 
corresponding dissipative fluxes also vanishes instantly all over the volume. 
Causality violation of dissipative hydrodynamics is corrected in 2nd order 
theories \cite{Israel:1976}.
According to this theory the shear and bulk viscous 
stress follows the relaxation equations,

\begin{eqnarray}
	D\Pi&=&\frac{1}{\zeta\beta_{0}}(\Pi+\zeta\theta)	\label{eq14}\\
D\pi^{\mu\nu}&=&\frac{1}{2\eta\beta_{2}}[2\eta\nabla^{<^{\mu}u^{\nu}>}-\pi^{\mu\nu}] \label{eq15}
\end{eqnarray}

\noindent where  $D=u^\mu \partial_\mu$ is the convective time derivative,
$\theta=\partial . u$ is the expansion scalar and $\nabla^{<\mu} u^{\nu>}= \frac{1}{2}(\nabla^\mu u^\nu + \nabla^\nu u^\mu)-\frac{1}{3}  
(\partial . u) (g^{\mu\nu}-u^\mu u^\nu)$ is a symmetric traceless tensor. One can identify $\zeta\beta_{0}$ and $2\eta\beta_{2}$ respectively with 
  the relaxation time for bulk and shear stress tensor,  $\tau_{\Pi}=\zeta\beta_{0}$ and $\tau_{\pi}=2\eta\beta_{2}$.

The relaxation equation \ref{eq14},\ref{eq15} for bulk and shear viscosity 
have additional terms according to the kinetic theory calculation \cite{arXiv:0907.2583}
and has the following form,
 
\begin{eqnarray}  
D\pi^{\mu\nu}&=&\frac{1}{\tau_\pi}[2\eta\nabla^{<^{\mu}u^{\nu}>}-\pi^{\mu\nu}] \nonumber   \\
&&-(u^\mu\pi^{\nu\lambda}+ u^\nu\pi^{\mu\lambda})Du_\lambda \label{eq16}\\
D\Pi&=&-\frac{1}{\tau_{\Pi}}[\Pi+\zeta\nabla_{\mu}u^{\mu}+\frac{1}{2}\zeta T\Pi\partial_{\mu}(\frac{\tau_{\Pi}u^{\mu}}{\zeta T})]. \label{eq17}
\end{eqnarray}

Assuming longitudinal boost invariance, earlier, we have solved the relaxation equations for shear stress tensors, using the code 'AZHYDRO-KOLKATA'. Details of the code can be found in \cite{Chaudhuri:2008sj}. We have extended the code to include the bulk viscosity. Relaxation equations for bulk and shear viscosity are solved simultaneously with the energy-momentum conservation equations. Details of the equations solved are given in the appendix~\ref{Appendix1}.

 \begin{figure}
	\centering
		\includegraphics[scale=0.5]{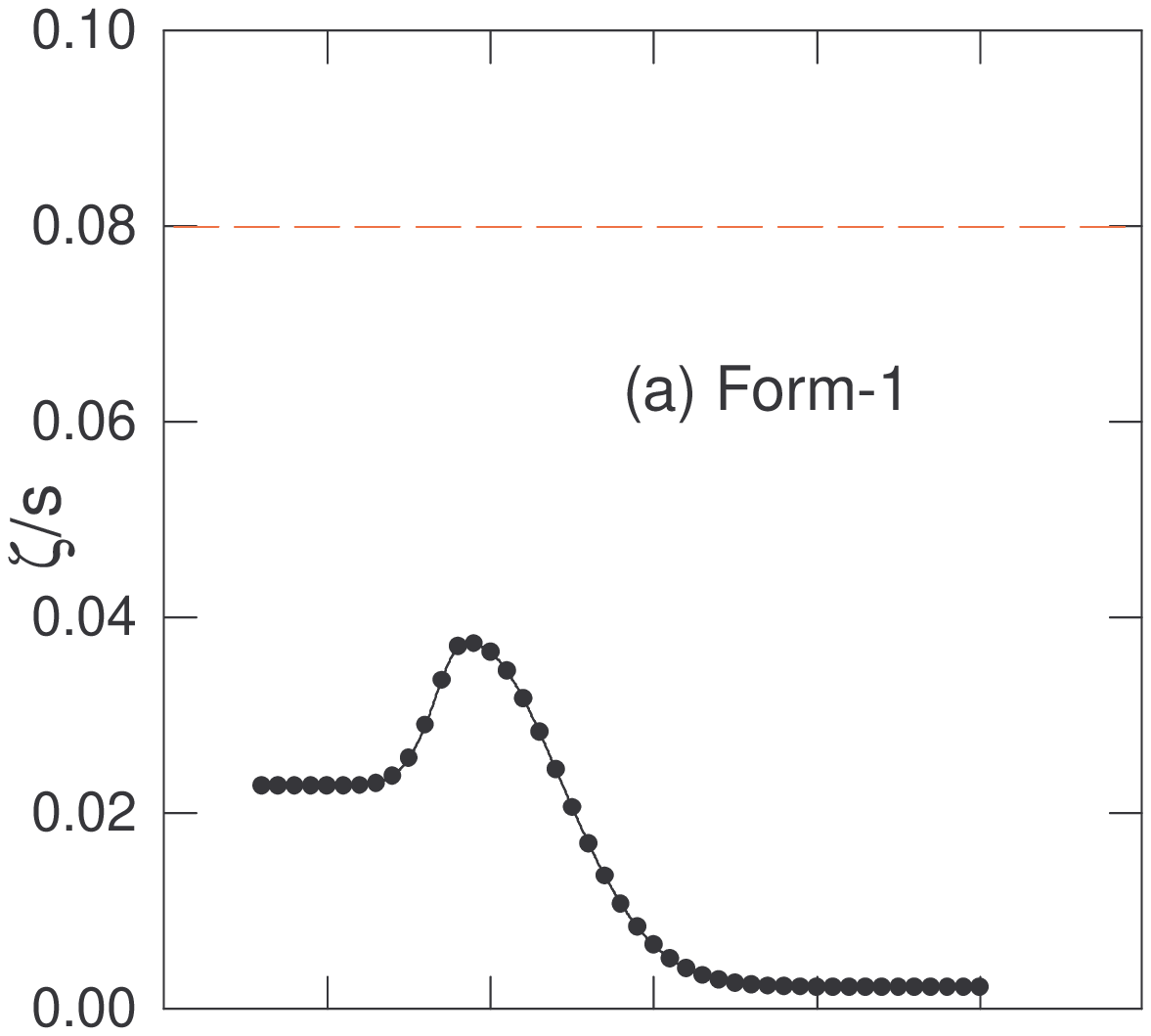}
		\includegraphics[scale=0.5]{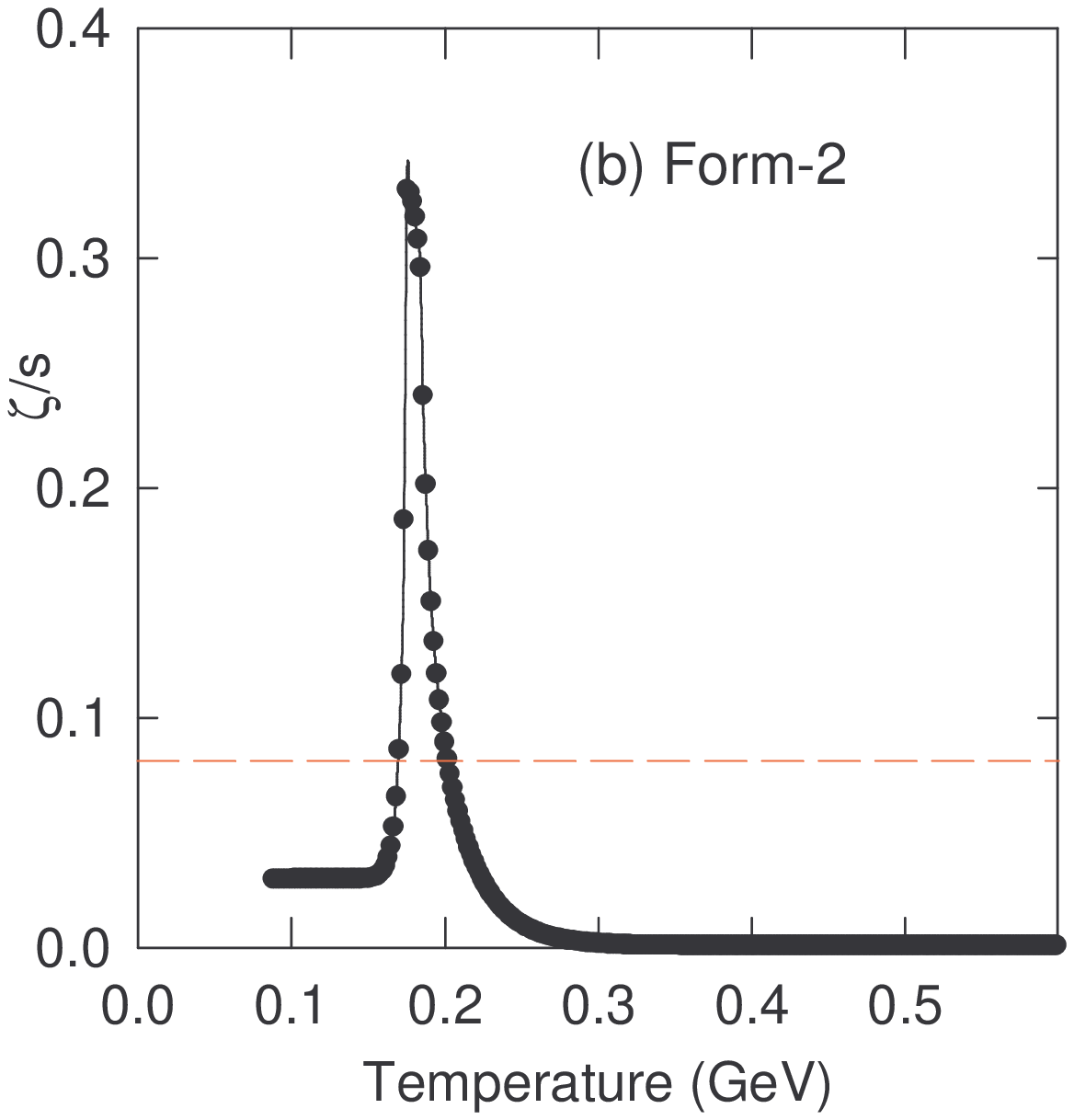}
	\caption{(Color online)Two different form of temperature dependence of $\zeta/s$.
	(a)Form-1: $\zeta/s$ in the QGP phase (T $>$175 MeV) is calculated by using PQCD formula
	  $\zeta/s=15\frac{\eta}{s}(T)(1/3-c^{2}_{s}(T))^{2}$,
	  where $c^{2}_{s}$ was calculated from recent lattice data \cite{Borsanyi:2010cj}. In the hadronic 
	  phase (T $<$ 175 MeV)$\zeta/s$ is parametrized from \cite{NoronhaHostler:2008ju}. 
	 (b)Form-2: This form is taken from \cite{Denicol:2009am}, where in the QGP phase
	  $\zeta/s$ was obtained from a different lattice calculation \cite{arXiv:0711.0914}
	 and $\zeta/s$ in hadronic phase is from \cite{NoronhaHostler:2008ju}. 
	 Red dashed line is the KSS bound \cite{Kovtun:2004de} of the shear viscosity
	 to entropy density ratio $\eta/s\sim1/4\pi$.}
	\label{fig:zeta}
\end{figure}

\begin{figure}
	\centering
	\includegraphics[scale=0.5]{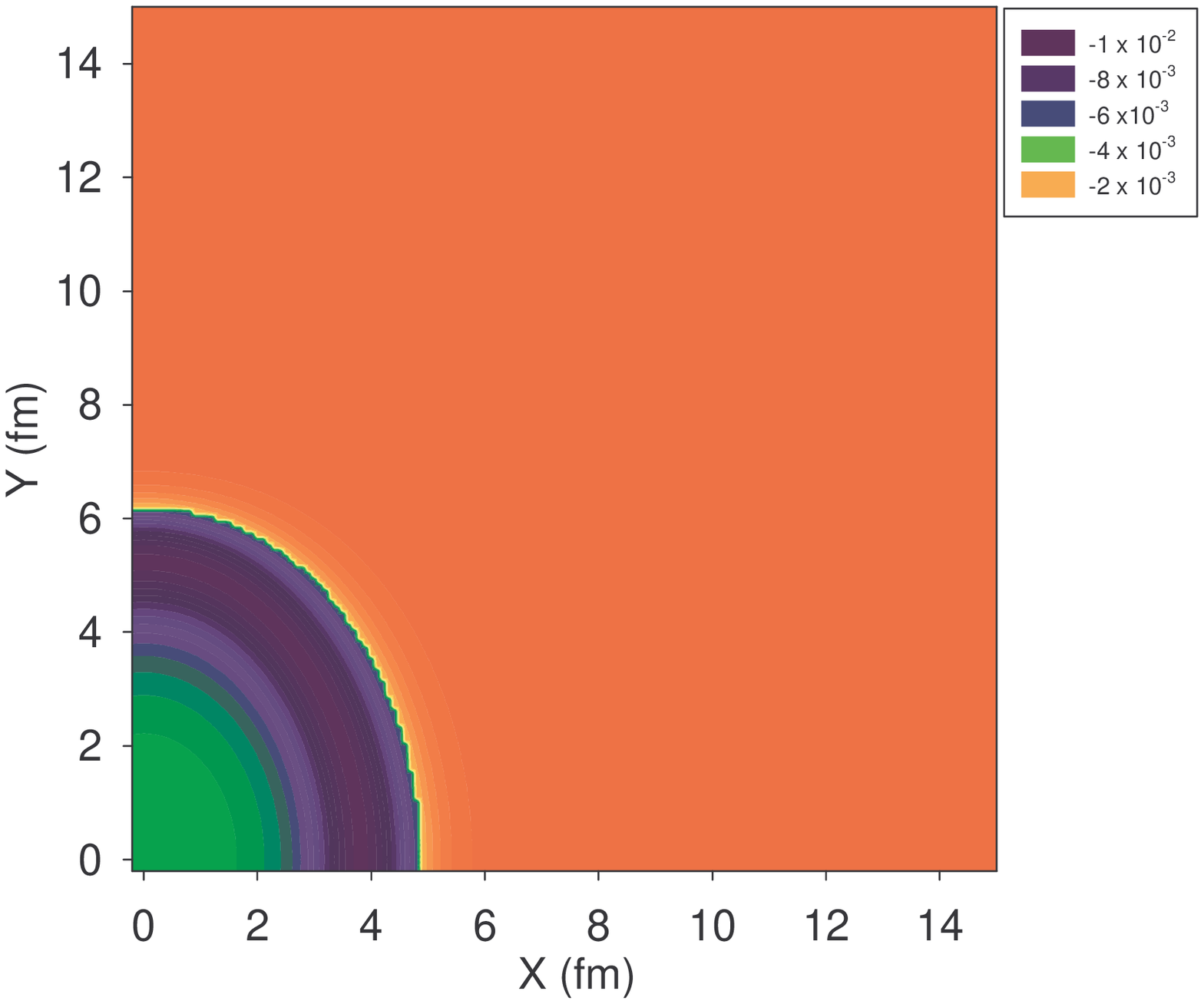}
	\includegraphics[scale=0.5]{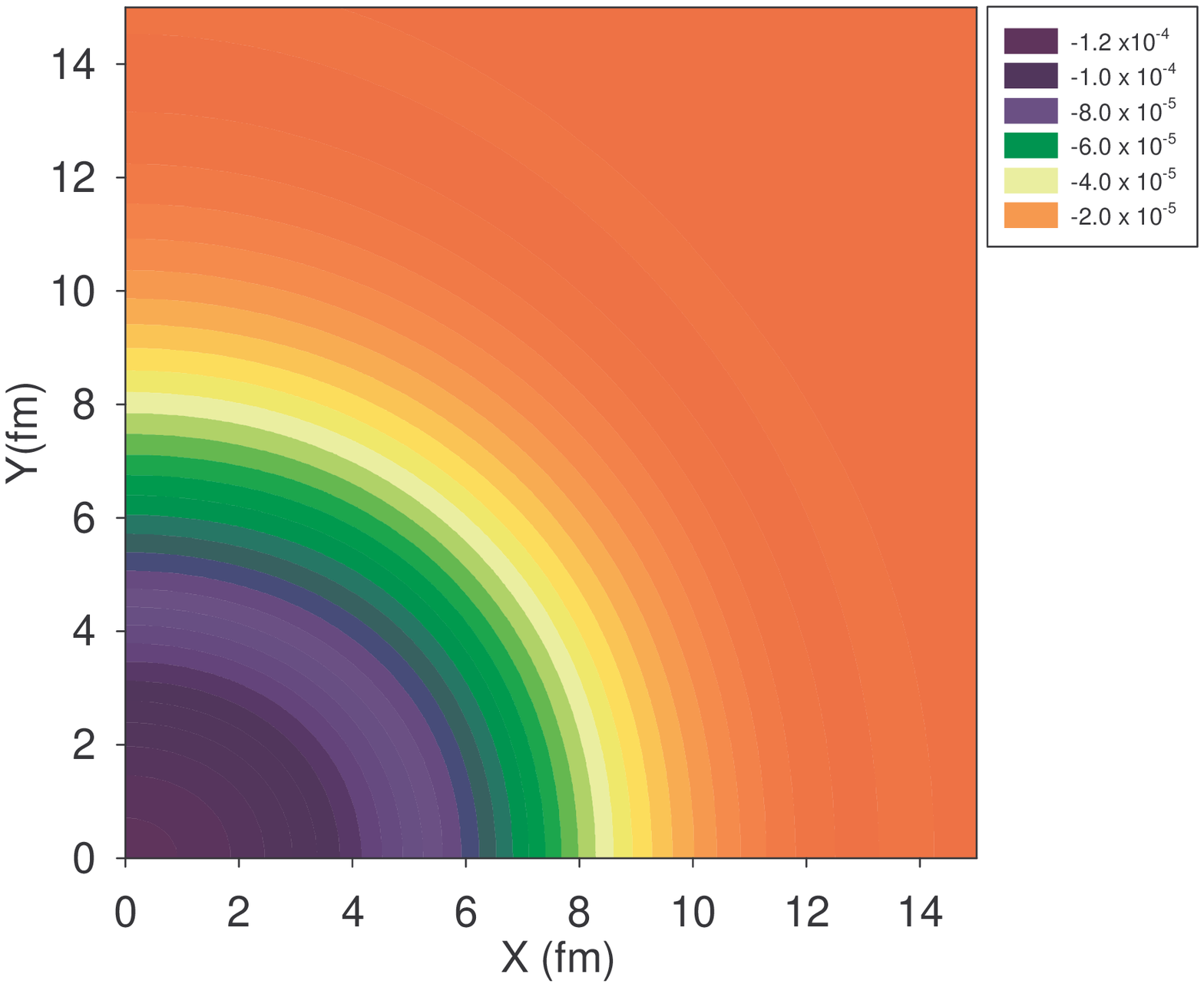}
	\caption{(Color online)Top Panel: the transverse profile of bulk stress 
	$\Pi(x,y)=-\zeta\theta$ at initial time $\tau_{0}$=0.6 $fm/c$. 
  Bottom Panel: $\Pi(x,y)$ at a later time $\tau=11.7 fm$.  }
	\label{fig:bulk}
\end{figure}

\begin{figure}
	\centering
		\includegraphics[scale=0.5]{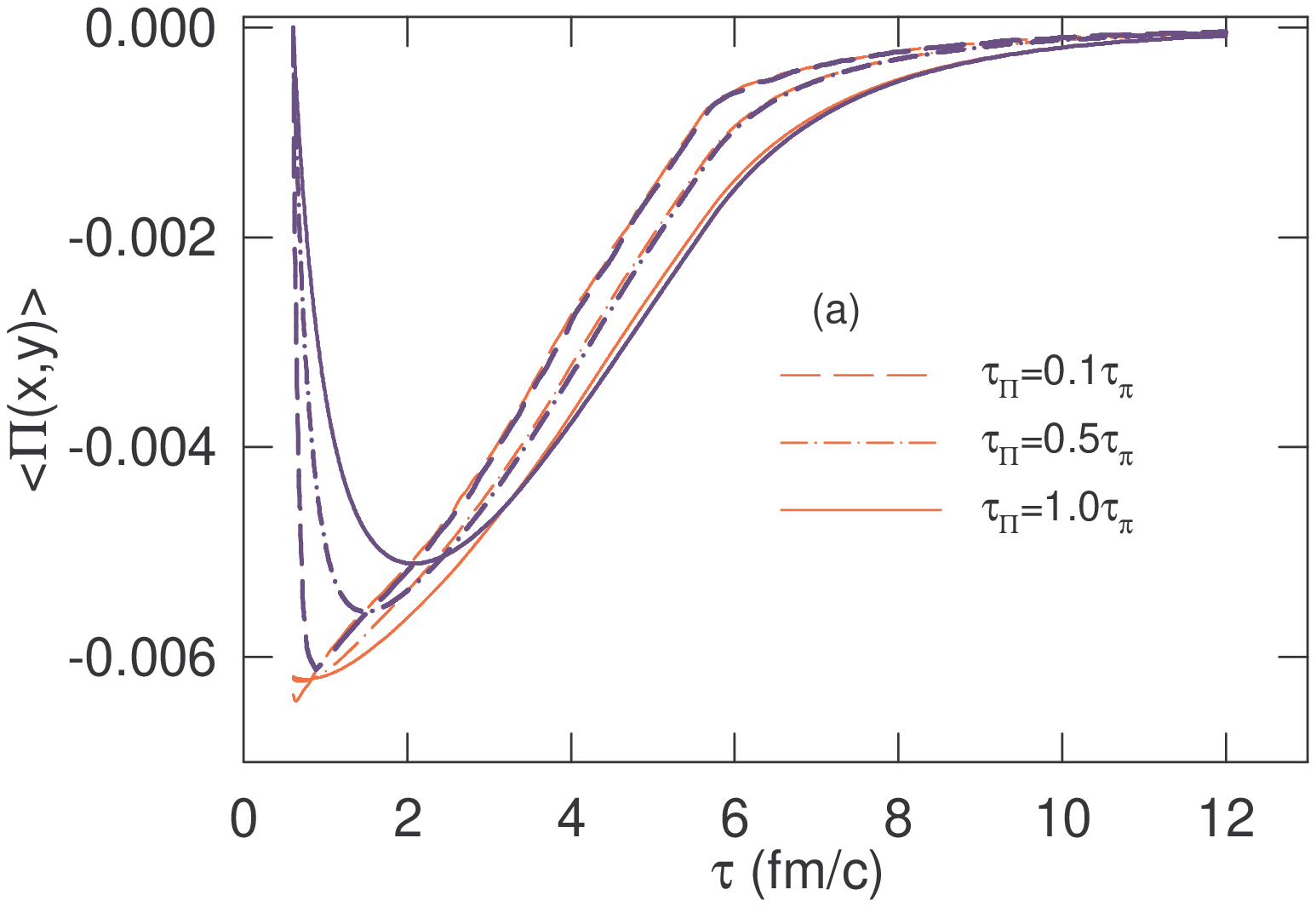}
		\includegraphics[scale=0.5]{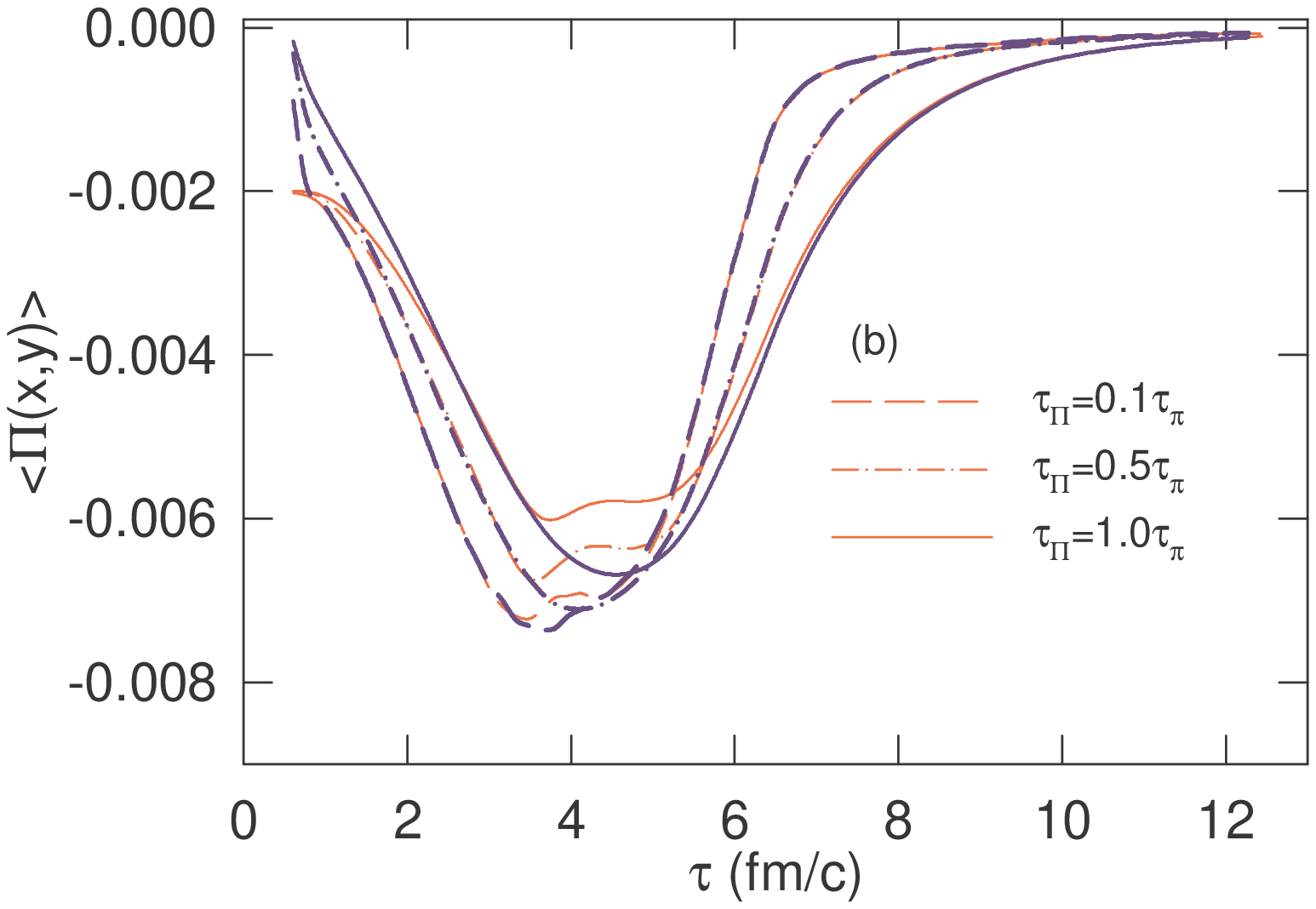}
	\caption{(Color online)The spatially averaged  bulk viscous stress $\Pi(x,y)$ in the 
	transverse plane as a function of evolution time for two different initialization
	of $\Pi(x,y)$and (a) for $\zeta/s$(T)form-1 (b)$\zeta/s$(T)form-2.  
  The simulation with zero initial bulk stress $\Pi(x,y)=0$ is marked
  as blue. Simulation for NS initialization $\Pi(x,y)=-\zeta\theta$ is marked as red. 
  The solid,dashed dotted
	and dashed lines represent the simulation with different relaxation time
	for $\Pi(x,y)$ namely $\tau_{\Pi}$=1.0,0.5and 0.1 times $\tau_{\pi}$.} 
	\label{fig:av_bulk}
\end{figure}

\section{Equation of state, viscosity coefficients and initial conditions}

\subsection{Equation of State}
Equation of state is one of the important input to hydrodynamic model. Through
this input, the macroscopic hydrodynamic models make contact with the microscopic
world. 
In the present simulations we have used an equation of state with cross-over transition at $T_c=175$MeV, developed earlier  \cite{Roy:2011xt}. The low temperature phase of the EoS is modeled by hadronic resonance gas, containing all the resonances with $M_{res} \leq$2.5 GeV. The high temperature phase is a parametrization of the 
recent lattice QCD calculation \cite{Borsanyi:2010cj}. Entropy density of the two phases were smoothly joined at $T=T_c=175 MeV$ by a step like function. 

  The thermodynamic
variable pressure,energy density etc. are then calculated by using the standard
thermodynamic relations.

\begin{eqnarray}
p\left(T\right)&=&\int^{T}_{0}s\left(T^{\prime}\right)dT^{\prime}\\
 \varepsilon\left(T\right)&=&TS\left(T\right)-P\left(T\right)
\end{eqnarray}

\subsection{Bulk and shear viscosity coefficients}
As discussed in the introduction the exact form of the  $\zeta/s$(T)
is quite uncertain. In this work 
we choose two different temperature dependent form. The form-1
is shown in Fig.\ref{fig:zeta}(a).
It is constructed in the following way, 
for the QGP phase we use the formula derived in pQCD calculation $\zeta/s=15\frac{\eta}{s}(T)\left(1/3-c^{2}_{s}(T)\right)^{2}$.
The squared sound speed $c^{2}_{s}(T)$ is calculated using the relation $c^{2}_{s}(T)=
\partial p(T)/\partial \epsilon (T)\mid_{s}$ where the pressure $p(T)$ and energy density 
$\epsilon (T)$ are obtained form the lattice QCD calculation \cite{Borsanyi:2010cj}. 
The $\eta/s$(T) used in this calculation is same as given in \cite{Niemi:2011ix} which was 
obtained using lattice data \cite{Nakamura:2004sy}. 
The $\zeta/s$ for the hadronic phase used in form-1 is the same as calculated in 
\cite{NoronhaHostler:2008ju}. Where it was calculated by using hadron resonance 
gas model including all known hadrons and their resonances up to mass 2 GeV with 
finite volume correction to thermodynamic quantities.
It also included an exponentially increasing density of Hagedorn states in the 
mass range of 2-80 GeV.    

The form-2 is shown in Fig.\ref{fig:zeta}(b). This form 
was taken from the reference \cite{Denicol:2009am}. For the QGP phase
the $\zeta/s$ was taken from a different lattice calculation \cite{arXiv:0711.0914}.
For the hadronic phase the $\zeta/s$ was taken from \cite{NoronhaHostler:2008ju}.

The peak value of form-2 is $\sim$10 times larger than the peak value in form-1.
Though both form of $\zeta/s$ shows a peak near the crossover temperature ($T_{co}\sim$175 MeV),
their dependency on temperature is slightly different in the QGP phase. 
The red dashed line in both Fig.\ref{fig:zeta}(a) and (b) shows the KSS bound of
shear viscosity to entropy density ratio\cite{Kovtun:2004de}.

For comparison purpose, we will also show some simulation result with only shear viscosity. However, in this demonstrative simulation, we have neglected the temperature dependence of shear viscosity to entropy ratio and perform the simulations with the AdS/CFT minimal value, $\eta/s=1/4\pi$.

\subsection{Relaxation Time}
Solution of relaxation equations for bulk and shear stress tensors requires to specify the relaxation time $(\tau_{\Pi})$ for the bulk stress and $(\tau_{\pi})$ for the shear stress tensor. In principle, relaxation times $\tau_\Pi$ and $\tau_\pi$ could be calculated from the underlying 
kinetic theory, which for  
strongly coupled QCD plasma, is a complex problem.   Relaxation times $\tau_\Pi$ and $\tau_\pi$ was
calculated in \cite{Israel:1976} for simple relativistic Boltzmann, Bose and Fermi gases
with mass $m$ using Grad 14 moment approximation in relativistic kinetic theory.
For a Boltzmann gas, in the non-relativistic limit ($\beta=\frac{m}{T} \rightarrow \infty$),   $\tau_{\Pi}=\zeta\beta_{0}\approx \frac{6}{5} \frac{m^2}{T^2}\frac{\zeta}{P}$ and  $\tau_\pi=2\eta\beta_2 \approx \eta/P$,
In the photon limit ($\beta \rightarrow 0$), $\tau_\Pi=\zeta \beta_{0}=\zeta \frac{216}{P}(\frac{kT}{m})^{4}$ and
$\tau_\pi=2\eta \beta_2=\frac{3\eta}{2P}$. Note that in the photon limit, the mass term appear in the the denominator with a quadratic power. The relaxation time become very large. For very large relaxation time, bulk stress will evolve very slowly with time. To circumvent the problem, for bulk stress also, one generally use the relaxation time for the shear stress tensor. 
In our simulation the $\tau_{\Pi}$ is either a constant or same as 
for the shear viscosity.

\subsection{Initial Conditions}

The initial conditions used here includes the initial energy density profile 
($\epsilon (x,y)$) in the transverse plane, the initial time ($\tau_{i}$)
,the transverse velocity profile $(v_{x}(x,y),v_{y}(x,y))$, shear and bulk stresses
in the transverse plane ($\pi(x,y),\Pi(x,y)$) respectively at $\tau_{i}$.
The values of these parameter are given in the table below.
One also need to specify the freezeout conditions to stop  the hydrodynamics 
evolution, this will be discussed in a later section.

\begin{table}[h]
\caption{\label{table1} Initial conditions for 2+1D viscous hydrodynamics calculation.} 
\begin{ruledtabular} 
  \begin{tabular}{|c|c|}
  Parameters                                 & Values  \\ \hline
$\epsilon_{0}$                               & 30 (GeV/$fm^{3}$)\\ \hline
$\tau_{i}$                                   &  0.6 fm  \\ \hline
$v_{x}(x,y),v_{y}(x,y)$                      & 0.0 \\ \hline 
$\pi^{xx}(x,y)=\pi^{yy}(x,y),\pi^{xy}(x,y)$  & $2\eta/3\tau_{i}$, 0  \\  \hline
$\Pi(x,y)$                                   & $-\zeta\theta$      \\  
 \end{tabular}\end{ruledtabular}  
\end{table}

The initial energy density profile in the transverse plane 
was parameterized in a two component Glauber model. At an impact
parameter $\bf{b}$, the transverse energy density is obtained
from the following two component form

\begin{equation} 
\epsilon({\bf b},x,y) = \epsilon_0[(1-x)N_{part} ({\bf b},x,y)+ x N_{coll}({\bf b},x,y)]
\label{eq:enprof}
\end{equation}  

\noindent where $N_{part}({\bf b},x,y)$ and $N_{coll}({\bf b},x,y)$ are the 
transverse profile of participant numbers and binary collision numbers 
respectively. $\epsilon_0$ in Eq.\ref{eq:enprof} does not depend on the impact parameter of the collision. It corresponds to the central energy density in b=0 impact parameter collision.
Generally $\epsilon_{0}$ is fixed to reproduce the experimental charged hadron multiplicity or the $p_T$ spectra in a central collision. Analysis of experimental data in $\sqrt{s}$=200 GeV Au+Au collisions indicate that the energy density of the central fluid is $\epsilon_0\approx$30 $GeV/fm^3$. In the present study, we fix $\epsilon_{0}=30 GeV/fm^{3}$. $x$ in Eq.\ref{eq:enprof} is the hard scattering fraction. It was shown in \cite{arXiv:0904.4080} that with x=0.13 the experimental centrality dependence of charged multiplicity data can be best explained at RHIC energy. The effect of varying 'x' has been studied in \cite{arXiv:1003.5791},\cite{hep-ph/0103234}. For the present study, we fix $x=0.13$. 
For the present study we carry out simulation for Au-Au collision at 
an impact parameter b=7.4 fm unless stated otherwise.

In this study we will use Navier-Stokes(NS) initialization
of bulk stress $\Pi(x,y)=-\zeta\theta$  unless stated otherwise.
One can also initialize $\Pi(x,y)$ by assuming a zero value at the
initial time $\tau_{i}$. 
Figure~\ref{fig:bulk} shows the initial $\Pi(x,y)$ in the
transverse plane with the Navier-Stokes initialization.




\subsection{Freezeout} 

In the following, we assume that the fluid undergoes kinetic freeze-out at a fixed temperature $T_F$. We have considered two freeze-out temperature, $T_F$=130 MeV and 160 MeV. In hydrodynamical model, one generally assumes that prior to kinetic freeze-out, fluid undergoes a 'chemical freeze-out' at   temperature $T_{chem} > T_F$, beyond which the particle ratios remain unaltered. In the present simulations, we have used an lattice QCD based EoS. Lattice QCD calculations are for zero baryon density fluid. We therefore assume that the chemical equilibration is maintained throughout the evolution.  The proton-antiproton  ratio will not be correctly reproduced in the model. However,
protons constitute only 5\% of the total charged particle yield. For other particles e.g. pion, kaon, the model simulation can correctly reproduce experimental numbers. Moreover, in the present simulations, we have studied the effect of bulk viscosity on hydrodynamic evolution and particle production. We have not attempted fits to experimental data.   

We have used the Cooper-Frye algorithm  \cite{CooperFrye} 
to evaluate particle spectra from the freeze-out surface. We have included the dissipative correction to particle spectra. For detail see the appendix~\ref{sec:Appendix2}.


 \begin{figure}
	\centering
		\includegraphics[scale=0.5]{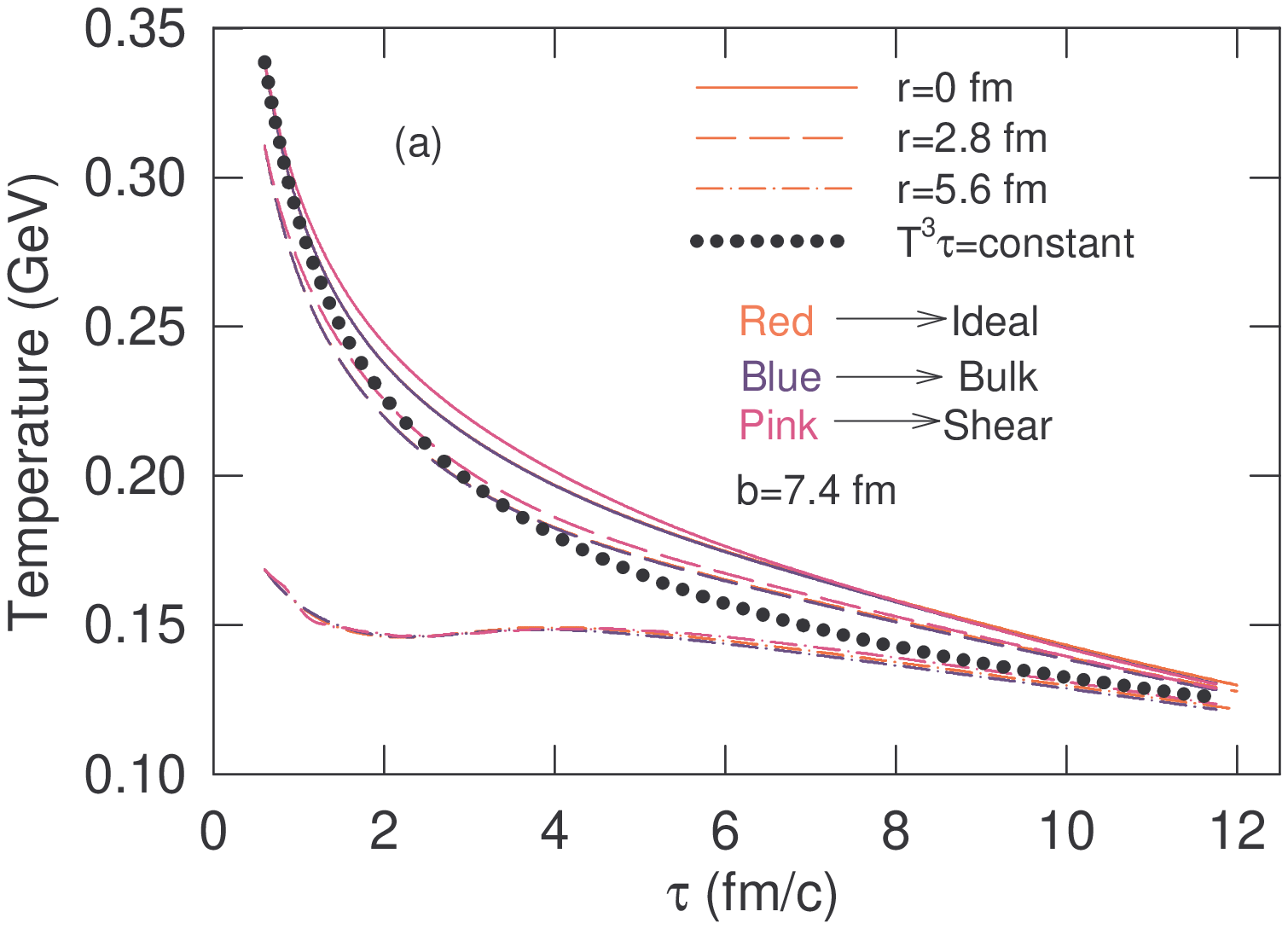}
		\includegraphics[scale=0.5]{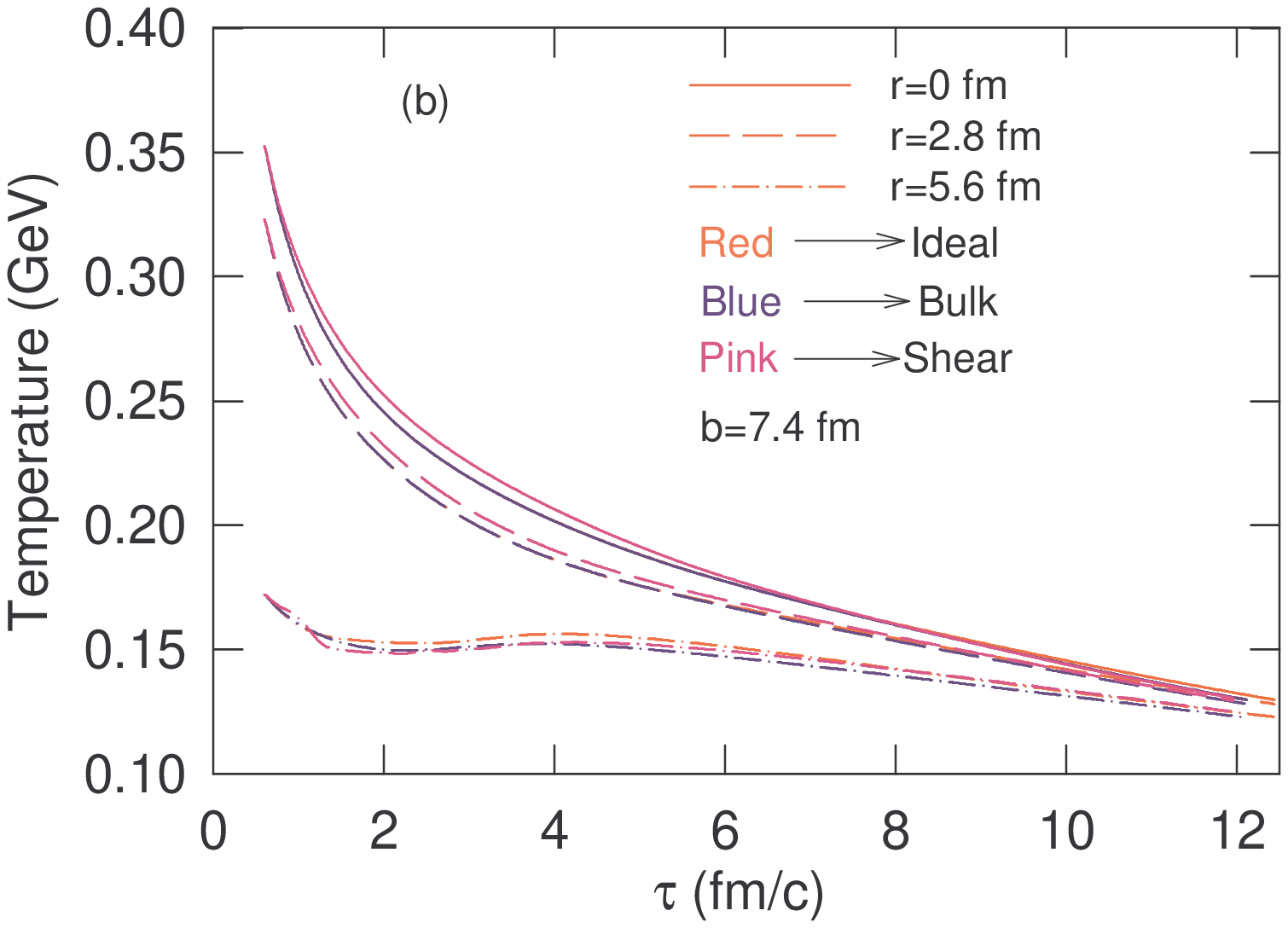}
	\caption{(Color online)The rate of cooling of the fluid for ideal, shear and bulk viscous evolution
	is shown here for three different location ,r=0 fm( solid line), at r=2.8 fm (dashed
	line) and at r=5.6 fm (dashed dotted line). Red line is for ideal fluid evolution, blue
	and pink lines are for bulk and shear viscous evolution respectively. Top panel (a) the
	calculation with $\zeta/s$ form-1. Bottom panel(b) with $\zeta/s$ form-2.
  }
	\label{fig:Temperature}
\end{figure}

\begin{figure}
	\centering
		\includegraphics[scale=0.5]{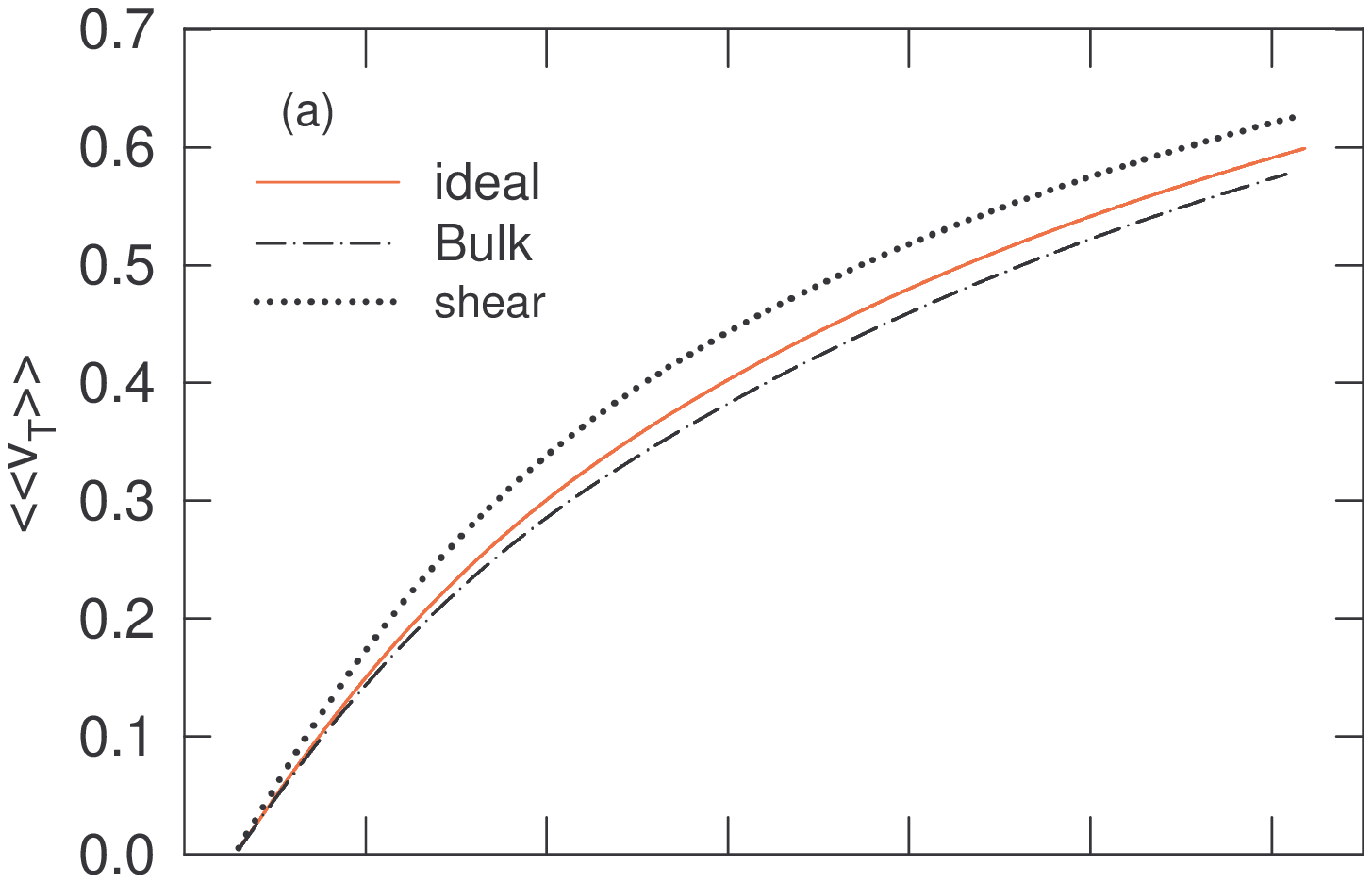}
		\includegraphics[scale=0.5]{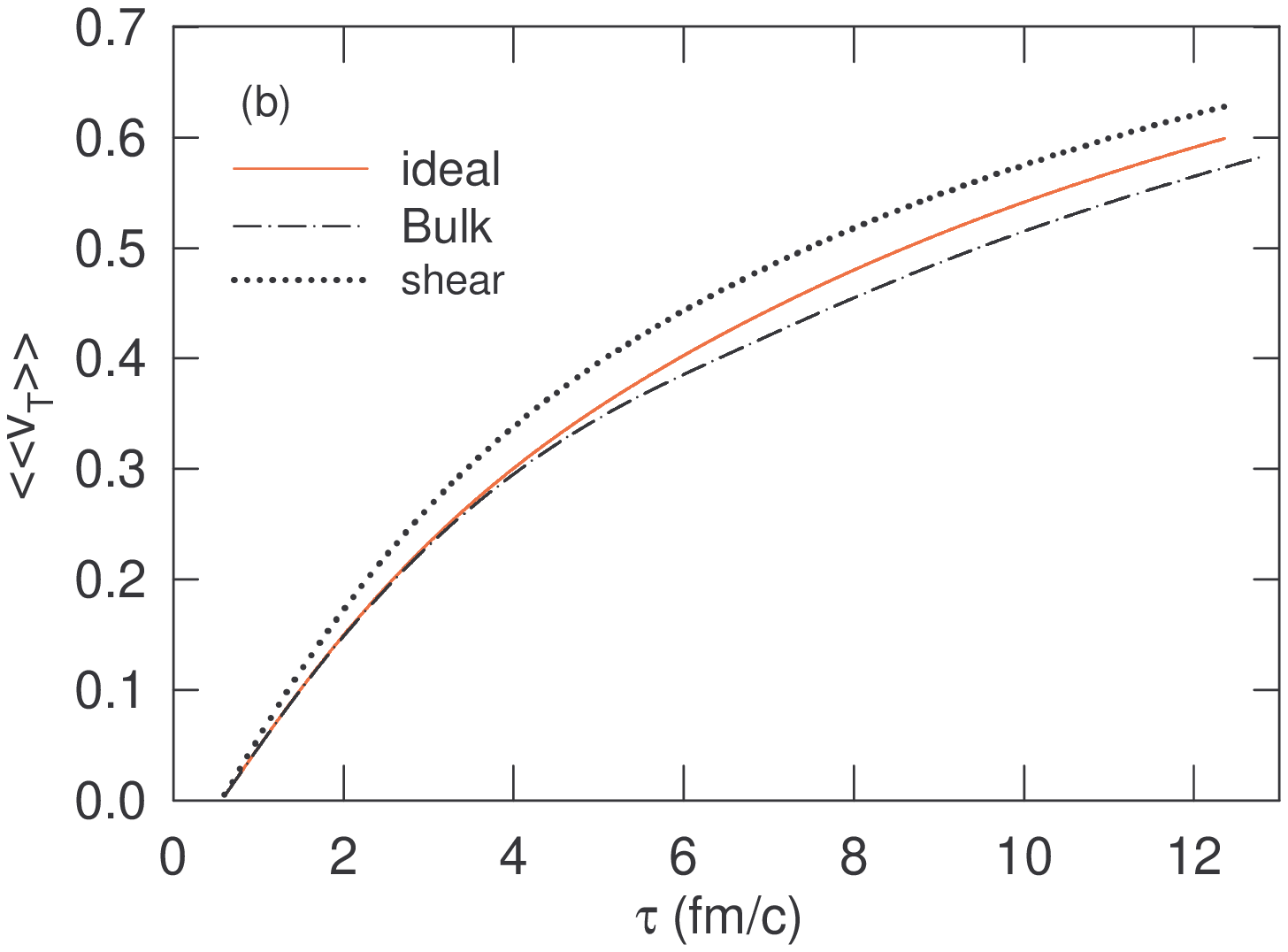}
	\caption{(Color online)Time evolution of spatially averaged transverse velocity $\left\langle v_{T}\right\rangle$
	for ideal and viscous simulation. The red solid line is for ideal fluid evolution
	and the dashed dot and dotted lines are for bulk and shear viscous evolution respectively.
	Top panel (a) is the simulation with form-1. Bottom panel (b) calculation with $\zeta/s$ form-2.}
	\label{fig:trans_vel}
\end{figure}

\section{Results}
\subsection{Effect of bulk viscosity on fluid evolution}
\label{sec:results}

The space-time evolution of $\Pi(x,y)$ is governed by the relaxation Eq.
\ref{eq17}.
The relaxation time in Eq.\ref{eq17} controls how fast the
stress $\Pi(x,y)$ relaxes to its instantaneous equilibrium value. 
The top panel of figure \ref{fig:bulk} shows the initial $\Pi(x,y)$ with 
Navier-Stokes initialization in b=7.4 fm Au+Au collisions for form-1 of bulk viscosity. In b=7.4 fm collision, collision zone is asymmetric. The bottom panel of the same figure 
shows the $\Pi(x,y)$ at a later time
$\tau$=11.76 fm. The relaxation time is $\tau_\Pi=\tau_\pi$.
Within a span of $\sim$10 fm, bulk stress is decreased by a factor of 100. One also note that with time, anisotropicity of the bulk stress is also decreased. Pressure gradient is more along the minor axis ($x$-direction in Fig.\ref{fig:bulk}), and fluid expands more in that direction reducing the anisotropicity. Form-II of $\zeta/s$ also give similar results. 

In Fig.~\ref{fig:av_bulk}(a) we have shown the temporal evolution of spatially 
averaged bulk stress, $\left\langle \Pi(x,y)\right\rangle$ for two different initializations (i) $\Pi(x,y)$=0 at the initial time $\tau_i$(blue lines) and
(ii) Navier-Stokes(NS) initialization $\Pi(x,y)=-\zeta\theta$
(red lines) at $\tau_i$. Results are shown for three different relaxation time, 
$\tau_{\Pi}$=1.0  (solid line),0.5  (dashed dot line) and 0.1  (dashed lines) times $\tau_{\pi}$. $\tau_{\pi}$=3$\eta/2p$ is the relaxation time for shear viscous stress estimated  for a relativistic Boltzmann gas\cite{Israel:1976}.
Late time evolution of bulk stress hardly depend on the initialization. Even when the bulk stress is initialized to zero value, it quickly reaches the Navier-Stokes value. Time by which the bulk stress reach the Navier-Stokes value depend on the relaxation time. The shorter relaxation time $\tau_{\Pi}$ means the system reaches its
equilibrium Navier-Stokes value faster, and in the limit of $\tau_{\Pi}\rightarrow 0$
Eq.\ref{eq17} transforms to relativistic Navier-Stokes equation (Eq.\ref{eq:NS}).  
From Fig.~\ref{fig:av_bulk} one can see that $\left\langle \Pi(x,y)\right\rangle$ with
zero initialization takes least time to attain its instantaneous equilibrium value (red line)
for the smallest  value of $\tau_{\Pi}$. In Fig.\ref{fig:av_bulk}(b), same results are shown for the form-2 of $\zeta/s$. 
Results are similar to that obtained for form-1 of $\zeta/s$. 


The rate of cooling of the fluid element for two different 
form of $\zeta/s$ at three different location in the reaction zone
for ideal (red line), bulk (blue line) and shear (pink line) viscous evolution are shown in 
Fig~\ref{fig:Temperature}(a) and (b).  Fig~\ref{fig:Temperature}(a) is the
simulation for $\zeta/s$ form-1 and ~\ref{fig:Temperature}(b) is for
$\zeta/s$ form-2.
The rate of cooling is different at various points in the
reaction zone. There is no noticeable change in the rate of cooling  due to  
bulk viscosity compared to ideal fluid evolution. For evolution with $\eta/s=0.08$ 
decreases the rate of cooling in the  early time for the central region.
The difference between the effect of bulk and shear stress on rate of cooling
is due to the difference in magnitude of $\Pi$ and $\pi^{\mu\nu}$. 
In the peripheral region the temperature variation is almost same for the ideal,
shear and bulk viscous evolution. Early in the evolution (2-3 fm) the fluid expansion
is mainly in the longitudinal direction and
follows Bjorken cooling law $T^{3}\tau=constant$. At a later time the transverse 
expansion leads to a different slope for the cooling rate.

\begin{figure}
	\centering
		\includegraphics[scale=0.5]{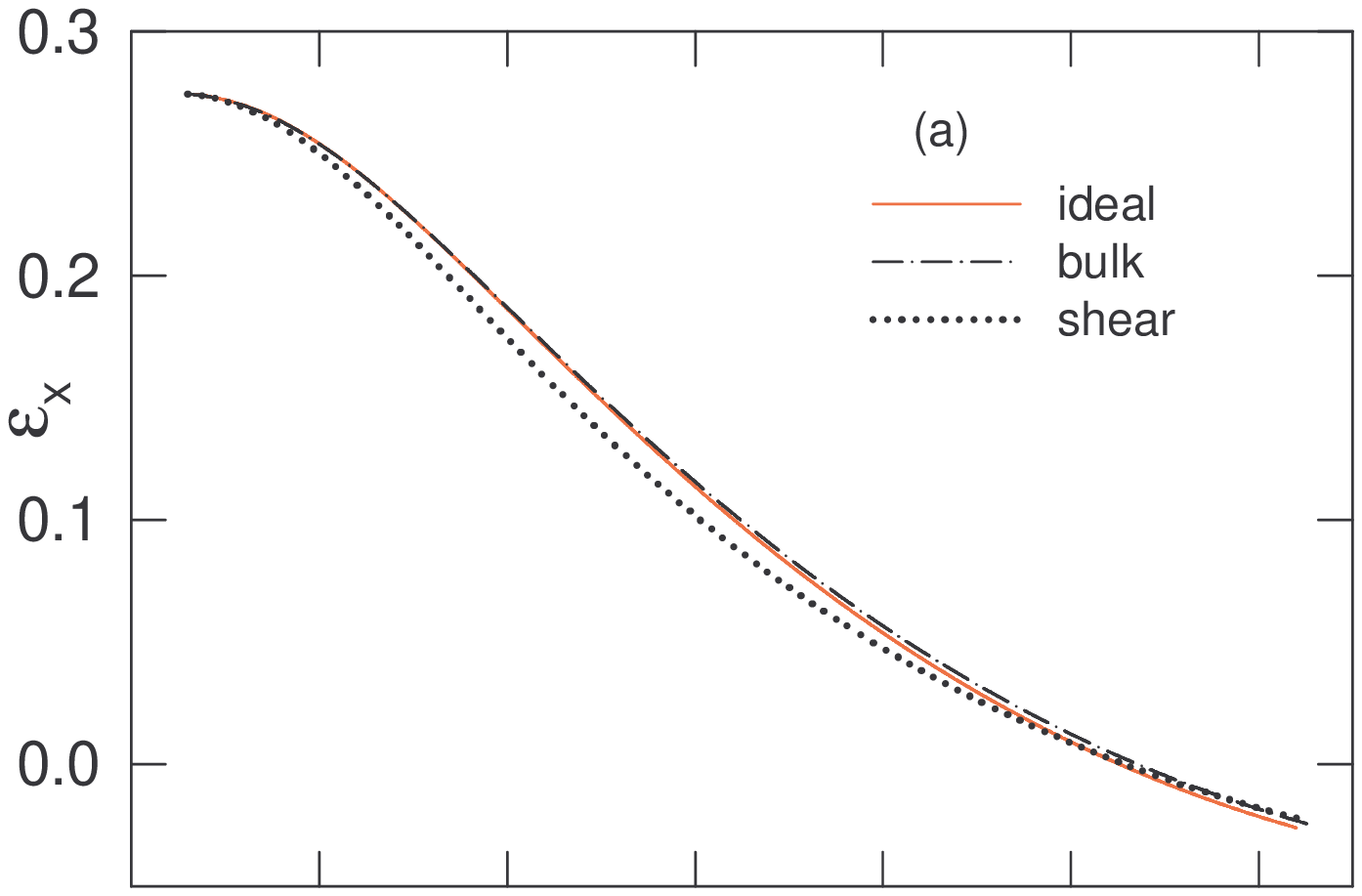}
	\includegraphics[scale=0.5]{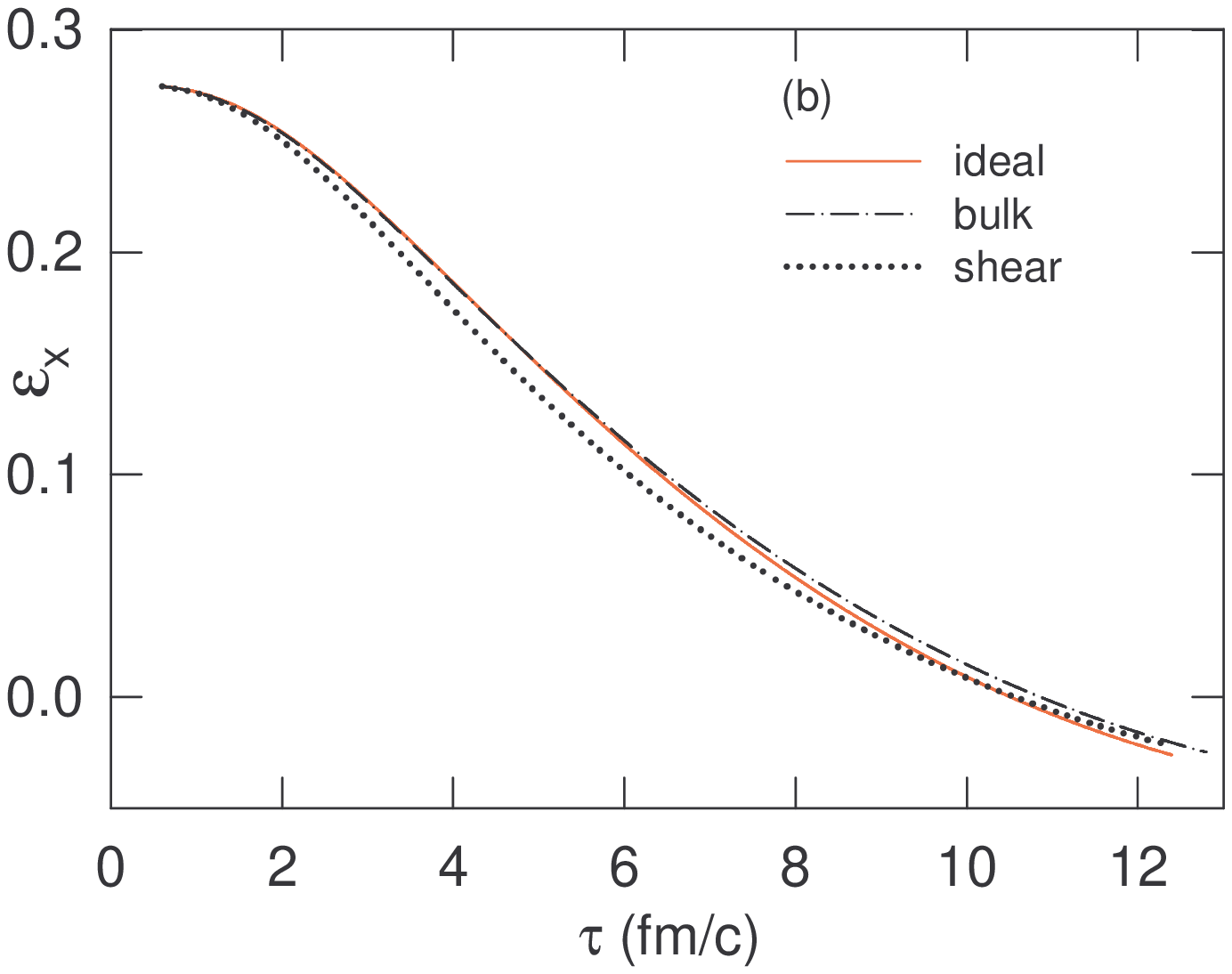}
	\caption{(Color online)The evolution of spatial anisotropy with time. Solid red line
	is the ideal evolution, dashed dotted line is for 
	bulk viscous and dashed line indicates the shear viscous evolution.
	(a) Simulation with $\zeta/s$ form-1. (b)Simulation with $\zeta/s$ form-2   }
	\label{fig:spat_ecen}
\end{figure}



Fig~\ref{fig:trans_vel}(a) and (b) shows the temporal  evolution of the spatially 
averaged transverse velocity ($\left\langle \left\langle v_{T}\right\rangle\right\rangle$)
of the fluid with form-1 and form-2 of 
$\zeta/s$ respectively.
The space averaged transverse velocity is defined as  
$\left\langle \left\langle v_{T}\right\rangle\right\rangle=\frac{\left\langle \left\langle \gamma\sqrt{v^{2}_{x}+v^{2}_{y}}~\right\rangle\right\rangle}{\left\langle \left\langle \gamma\right\rangle\right\rangle}$.
Solid red line is for ideal fluid and the dashed dotted and
dotted line is for bulk and shear viscous evolution respectively.
Because of the reduced  pressure gradient in the  bulk viscous evolution
compared to ideal fluid, the corresponding 
$\left\langle \left\langle v_{T}\right\rangle\right\rangle$ gets reduced.
The reduction in $\left\langle \left\langle v_{T}\right\rangle\right\rangle$
at later time (after $\sim$ 8 fm) is more for the $\zeta/s$ form-2 compared to 
form-1.
The shear viscosity on the other hand increases the pressure gradient in the
transverse direction and reduced the longitudinal pressure at the early time of evolution.
Because of the enhanced pressure gradient ,the $\left\langle \left\langle v_{T}\right\rangle\right\rangle$
for shear viscous evolution is increased compared to ideal fluid.\\

\begin{figure}
	\centering
		\includegraphics[scale=0.5]{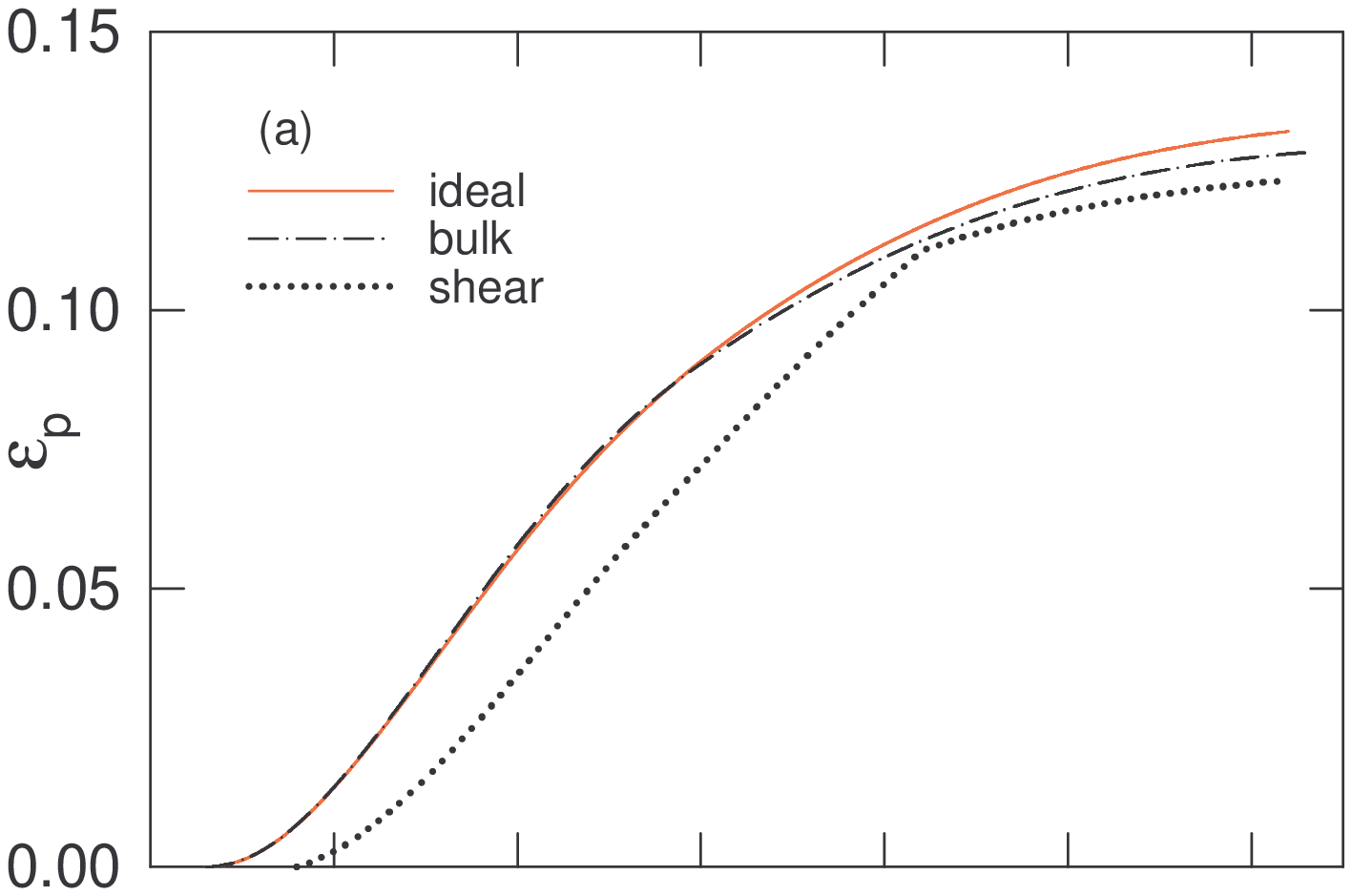}
		\includegraphics[scale=0.5]{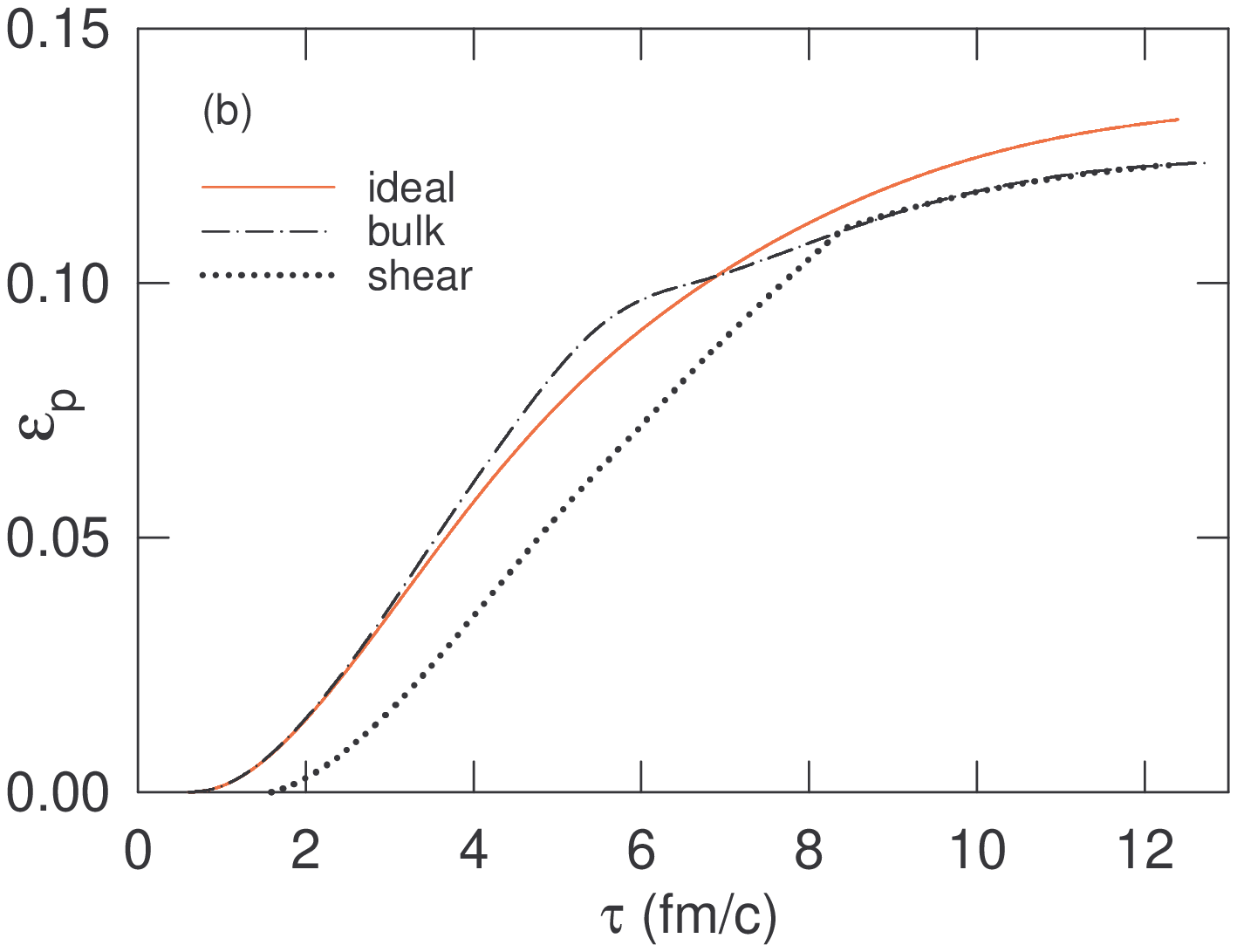}
	\caption{(Color online)The evolution of momentum anisotropy with time. The lines shows 
	here are the same as explained in figure~\ref{fig:spat_ecen}. Fig.(a) is the result for form-1 of
	$\zeta/s$ and (b) is for form-2 of $\zeta/s$.}
	\label{fig:mom_ani}
\end{figure}

In a non-central collision between two identical nuclei, the collision zone is non-spherical. The spatial eccentricity $\varepsilon_{x}$ is defined as 

\begin{equation} \label{eq26}
	\varepsilon_{x}=\frac{\left\langle \left\langle y^{2}-x^{2}\right\rangle\right\rangle}{\left\langle \left\langle y^{2}+x^{2}\right\rangle\right\rangle},
\end{equation}

\noindent is a measure of spatial deformation of the fireball from spherical shape.  
A zero 
value of $\varepsilon_{x}$ means the system is spherical, $0<\varepsilon_{x}<1$ indicates
elliptic shape with major axis along Y direction, and $\varepsilon_{x}<0$ means
the major axis along X direction.
The angular bracket $\left\langle \left\langle...\right\rangle\right\rangle$ 
implies an energy density weighted average.
For b=7.4 fm collision,the evolution of $\varepsilon_{x}$ with time 
($\tau$) is depicted in Fig.~\ref{fig:spat_ecen}(a) and (b). 
 The solid red line corresponds to the 
temporal evolution of $\varepsilon_{x}$ for ideal fluid, the  dashed
dotted and dashed line are for bulk and shear viscous fluid evolution respectively.
Because of the  reduced
pressure gradient in bulk viscous evolution, 
the initial spatial deformation  ($\varepsilon_{x}\approx 0.28$)
takes a longer time to change its
shape  for the bulk viscous evolution  compared to the
ideal fluid evolution. At a later time the change in the $\varepsilon_{x}$
is more pronounced for form-2 of $\zeta/s$ compared to form-1.
Shear viscosity does the opposite to the transverse 
expansion, initially it enhances the transverse velocity
and the spatial deformation $\varepsilon_{x}$ reduces at a much higher
rate compared to the ideal fluid evolution. \\
\begin{figure}
	\centering
		\includegraphics[scale=0.5]{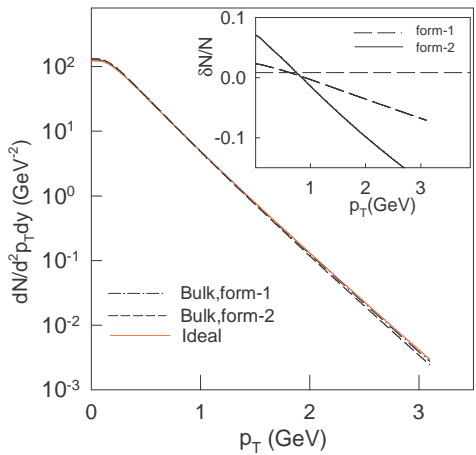}
	\caption{(Color online)Negative charged pion's $p_{T}$ spectra for Au+Au collision at impact parameter
	b=7.4 fm.The solid red line is the spectra from ideal fluid. The dashed dot and dotted lines are for   fluid with bulk viscosity with form-1 and form-2. 
	The inset figure shows the ratio of correction to the $p_{T}$ spectra due to 
	bulk viscosity to ideal evolution.}
	\label{fig:combined_pt_spectra}
\end{figure}

Similar to the spatial anisotropy, one can define the asymmetry of
fireball in momentum space. The momentum space anisotropy $\varepsilon_{p}$
is defined as  
\begin{equation}
	\varepsilon_{p}=\frac{\int dxdy (T^{xx}-T^{yy})}{\int dxdy (T^{xx}+T^{yy})}
 \end{equation}

The simulated elliptic flow $v_{2}$ in hydrodynamic model is directly related to the temporal evolution of the momentum anisotropy. In fact in ideal hydrodynamics $v_{2}\propto\varepsilon_{p}$. The evolution of $\varepsilon_{p}$ as a function of time is shown in Fig.~\ref{fig:mom_ani}(a) and (b) for $\zeta/s$ form-1 and $\zeta/s$ form-2 respectively. Compared to ideal evolution (solid red line), the bulk viscous evolution (dashed dot line) results in a reduced value of momentum anisotropy around freezeout time($\sim$ 12 fm/$c$). The change in $\epsilon_{p}$ for bulk viscous evolution compared to ideal fluid occurs after $\tau \sim$ 3-4 fm/$c$.  Around this time most regions of the fluid element reaches the temperature range $\sim$ 175 MeV where $\zeta/s$ has maximum value.\\ 

\begin{figure}
	\centering
		\includegraphics[scale=0.5]{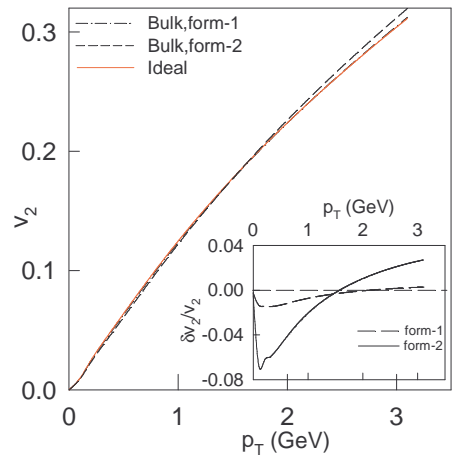}
	\caption{(Color online) Negative charged pion's $v_{2}$ for Au+Au collision at impact parameter
	b=7.4 fm.The solid red line is the $v_{2}$ from ideal fluid. The dashed dot and dotted lines are for
	   fluid with bulk viscosity with form-1 and form-2. 
	The inset figure shows the ratio of correction to the $v_{2}$ due to 
	bulk viscosity to ideal evolution.}
	\label{fig:combined_v2}
\end{figure}

\subsection{Effect of bulk viscosity on particle spectra and elliptic flow} 
\subsubsection{Without freeze-out correction}

We first discuss the change in particle spectra and elliptic flow due to bulk viscosity. Non-equilibrium correction to equilibrium distribution function is neglected.  The simulated charged pion $p_{T}$ spectra is shown in the Fig.~\ref{fig:combined_pt_spectra}. The red solid line in \ref{fig:combined_pt_spectra}  is the result obtained in ideal fluid simulation, the dashed dot and dotted line is obtained with form-1 and form-2 of bulk viscosity respectively.  
Due to the smaller transverse flow in presence of bulk viscosity  we get a steeper $p_{T}$ spectra compared to the ideal fluid evolution. However, the change is small. The inset figure shows the relative correction in the $p_{T}$ spectra $\delta N/N_{eq}$ due to the bulk viscosity  compared to ideal simulation. Where $\delta N$= $N_{bulk}-N_{eq}$. The correction to the $p_T$ spectra is small, at $p_{T}\sim$2 GeV the correction is $\sim$ 10\% with form-2 of $\zeta/s$. It is even less for form-1.  
The simulated elliptic flow $v_{2}$ of $\pi^{-}$ produced in Au-Au collision at impact parameter b=7.4 fm collision are shown in Fig.~\ref{fig:combined_v2}. The red solid line is the result for ideal hydrodynamics. The dashed dotted and dotted  lines are $v_2$ with bulk viscosity with form-1 and form-2 respectively. Here, again, we have not included the non-equilibrium correction to the distribution function.  The inset of ~\ref{fig:combined_v2} shows the relative change in the $v_{2}$ with bulk viscosity compared to ideal fluid evolution. The relative correction in $v_{2}$ due to the form-1 of $\zeta/s$ is within $\sim$ 4\% for the $p_{T}$ range 0-3 GeV and for form-2 the relative correction is within $\sim$ 8\%. It appear that if the non-equilibrium correction to the equilibrium distribution function is neglected,  both the form of bulk viscosity introduces relatively small correction to the particle spectra and elliptic flow. 

\begin{figure}
	\centering
		\includegraphics[scale=0.5]{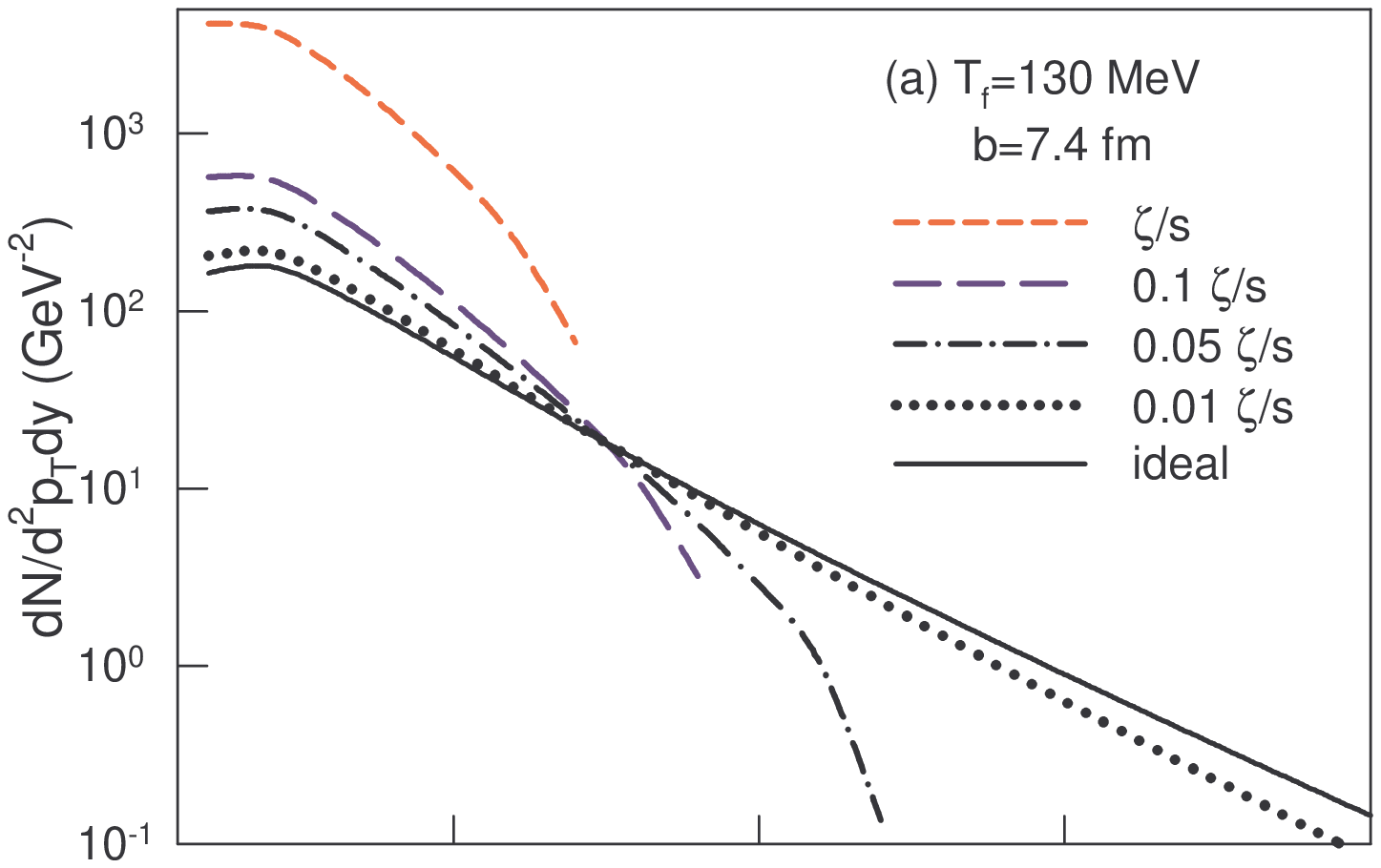}
  	\includegraphics[scale=0.5]{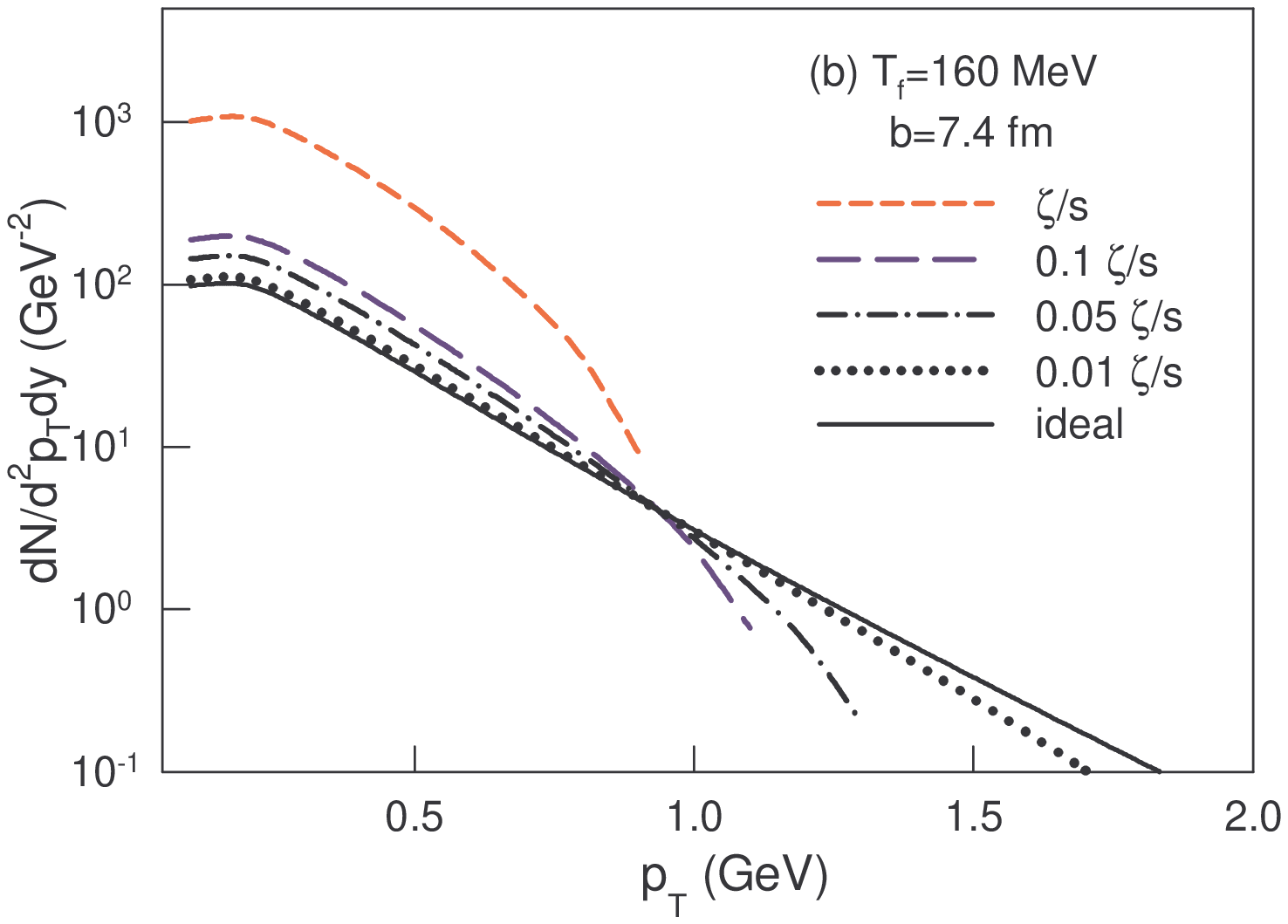}
	\caption{(Color online) The charged pion $p_{T}$ spectra for 
	impact parameter b=7.4fm Au-Au collision. Ideal fluid evolution
	(black solid line) and various values of $\zeta/s$ form-1. (a) Freezeout
	temperature $T_{f}$=130 MeV (b) $T_{f}$=160 MeV.}
	\label{fig:grad_spectra}
\end{figure}

 \begin{figure}
	\centering
		\includegraphics[scale=0.5]{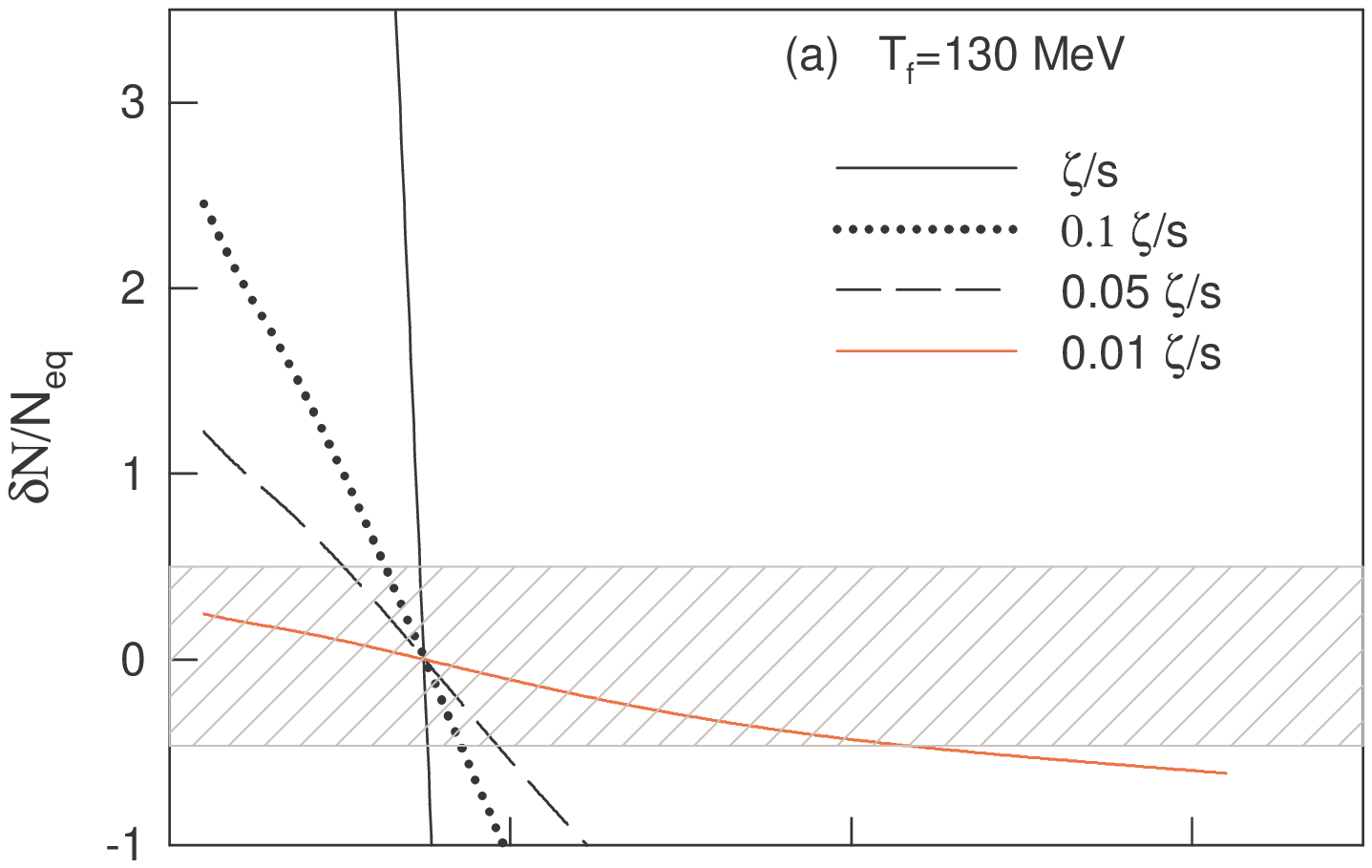}
		\includegraphics[scale=0.5]{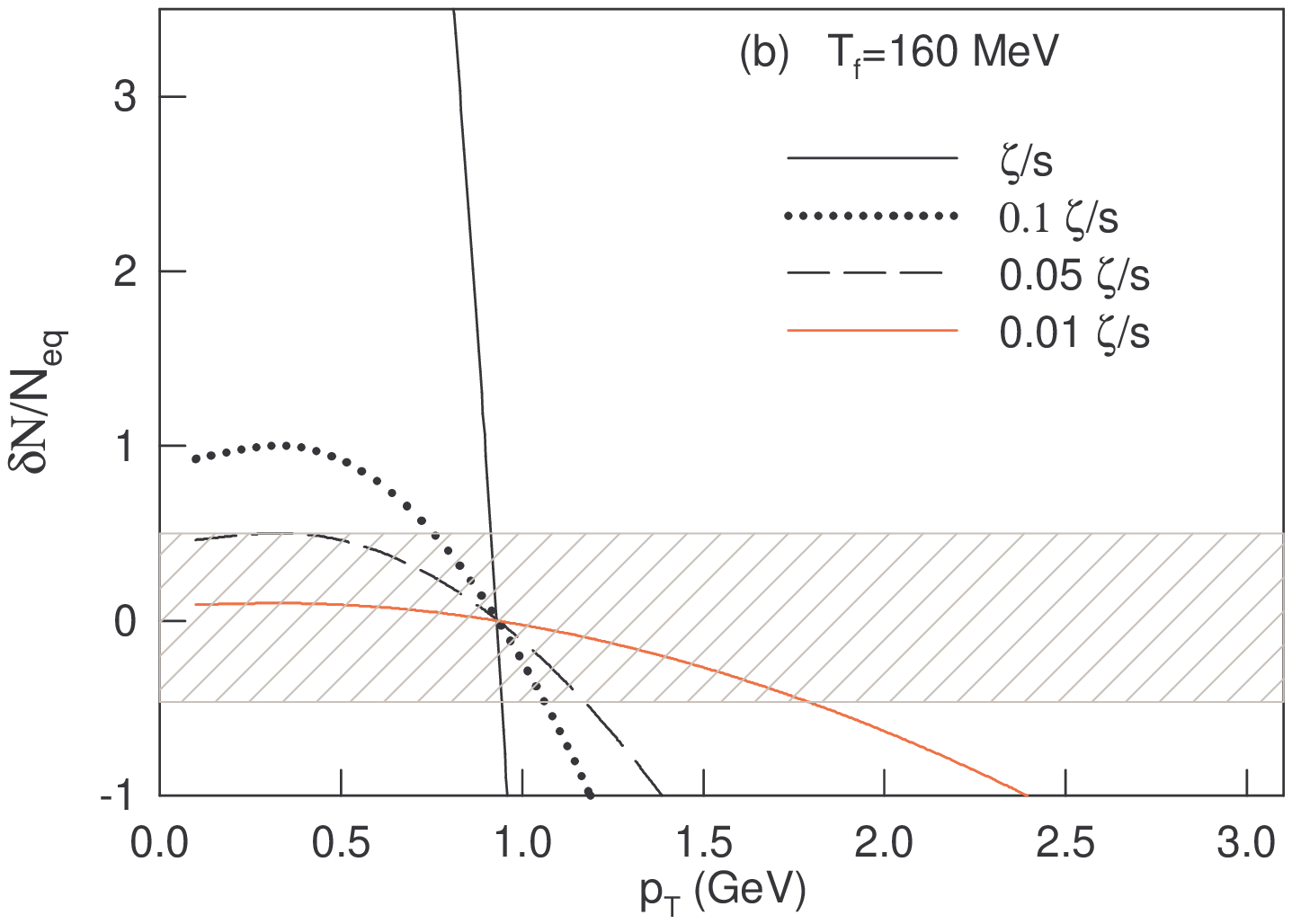}
	\caption{(Color online)The relative correction to the $p_{T}$ spectra due to the bulk viscosity
	compared to ideal fluid evolution. Dissipative correction in both evolution as well 
	in freezeout distribution	function has been included.
	Shaded region shows the relative correction 
	of 50\%.(a) Freezeout temperature $T_{f}$=130 MeV (b) $T_{f}$=160 MeV.   
	}
	\label{fig:grad_130}
\end{figure}

\subsubsection{With freeze-out correction}
\label{sec:freezeoutCorrection}

It has already been discussed that there are two kinds of dissipative correction to 
the ideal fluid simulation. First the energy momentum
tensor contains a viscous correction and the freezeout distribution function 
is also modified in the presence of dissipative processes. 
So far in this paper all the bulk viscous simulation results are obtained for 
dissipative correction to the energy momentum tensor only.
In this section we discuss the correction to the freezeout distribution function.
 We have employed Grad's fourteen-moment methods
for freezeout dissipative correction as described in~\cite{Monnai:2009ad}.
The implementation of this method to our 2+1D viscous code "`AZHYDRO-KOLKATA"'
is briefly discussed in the appendix~\ref{sec:Appendix2}.

Figure \ref{fig:grad_spectra} (a) and (b) shows the $p_{T}$ spectra of pion
for freezeout temperature $T_{f}$=130 and 160 MeV for four different values
of $\zeta/s$ form-1. 
The black solid line in Fig.~\ref{fig:grad_spectra} is the $p_{T}$ spectra for 
ideal fluid evolution in 20-30\% Au-Au collision, the other lines are for bulk  
viscous evolution with varying values of form-1 $\zeta/s$.
The red dashed line is the simulated spectra with form-1 $\zeta/s$, 
whereas long dashed, dashed dotted and dotted lines are results for 0.1,0.05 and 
0.01 times the form-1 $\zeta/s$.  It appears that for QCD motivated bulk viscosity over entropy ratio, $(\frac{\zeta}{s})_{QCD}=15\frac{\eta}{s}(\frac{1}{3}-c_s^2)^2$,  Grad's14-moment method introduces very large correction to the spectra. At low $p_T$, compared to ideal fluid, $\pi^-$ yield is increased by a factor of 10 or more. Corrections are comparatively less if bulk viscosity is reduced. For very small bulk viscosity, 0.01 times form-1 $\zeta/s$, the spectra is very close to the ideal fluid evolution in the $p_{T}$ range $0<p_{T}<1 GeV$.

To put things in perspective for the present calculations relative to the 
existing calculations in \cite{Monnai:2009ad}, we should compare the relative 
change in invariant yield of negative pions versus $p_{T}$ between ideal 
simulation and bulk viscous simulation for both cases. In order to do that, 
we have to compare with the corresponding calculations done in the present 
work with input bulk viscosity to entropy density of $0.01\times\zeta/s$ at 
$T_{f}$ = 160 MeV.  Such a value of $\zeta/s$ is chosen so as to have a 
similar magnitude of $\Pi$ over the freeze-out surface as used in \cite{Monnai:2009ad}.
We find that the relative corrections are similar.

The specific form of the dissipative correction to the ideal freezeout distribution
function considered here (see appendix~\ref{sec:Appendix2}) leads to a large negative correction
to the $p_{T}$ spectra for large values of $p_{T}$. Depending on the value of $\zeta/s$ the 
dissipative correction due to the bulk viscosity results in a negative invariant yield above 
a certain value of $p_{T}$.
A negative value of particle number is unphysical, we will omit the $p_{T}$ range
in the subsequent plots from where the particle number becomes negative.
As discussed earlier, freeze-out correction is obtained under the assumption that the non-equilibrium correction to the distribution function is small than the equilibrium distribution function,  $\delta f_{bulk} << f_{eq}$. It is then implied that the relative correction ($\delta N/N_{eq}$) is small for Israel-Stewart's hydrodynamics to be applicable. Figure \ref{fig:grad_130}~(a) and (b) shows the corresponding
relative correction ($\delta N/N_{eq}$)
to the $p_{T}$ spectra due to the bulk viscosity. 
The shaded band in the figure corresponds to the relative correction of 50\%.
We consider here a correction of magnitude greater than 50\% 
indicates the breakdown of the the freezeout correction procedure.  
 From Fig.~\ref{fig:grad_130} one can see that 
the $p_{T}$ spectra changes drastically in bulk viscous evolution (solid black line)
with $\zeta/s$ form-1. Only for viscous simulation with bulk viscosity to entropy
density ratio less than 0.01 times $\zeta/s$ the relative correction is less than 50\%. 
The results impose a severe constraint on the bulk viscosity, 
$\frac{\zeta}{s} \leq 0.01 (\frac{\zeta}{s})_{QCD}$.

We find from Fig.~\ref{fig:grad_spectra} and \ref{fig:grad_130} that the value 
of $\delta f_{bulk}$ 
is greater for the simulations with $T_{f}$ = 130 MeV than those 
for $T_{f}$ = 160 MeV. This can be understood in the following manner.
As shown in the last equation of Appendix~\ref{sec:Appendix2} both the values of $\Pi$ 
on the freeze-out surface as well as the values of the prefactors 
$D_{0}$, $B_{0}$ and $\tilde{B}_{0}$ at a given value of $T_{f}$ 
determines the magnitude of $\delta f_{bulk}$.  
The average magnitude of $\Pi$ decreases from about 
$-5 \times 10^{-6} GeV/fm^{3}$ at $T_{f}$ = 160 MeV to 
$-2\times 10^{-5} GeV/fm^{3}$ at $T_{f}$ = 130 MeV. 
However the prefactor values as given in Table~\ref{table2} 
increases with decrease in temperature. So the observed $T_{f}$ 
dependence of $\delta f_{bulk}$ is due to 
the interplay of both the temperature dependence of $\Pi$ and the prefactors.

\begin{figure}
	\centering
		\includegraphics[scale=0.5]{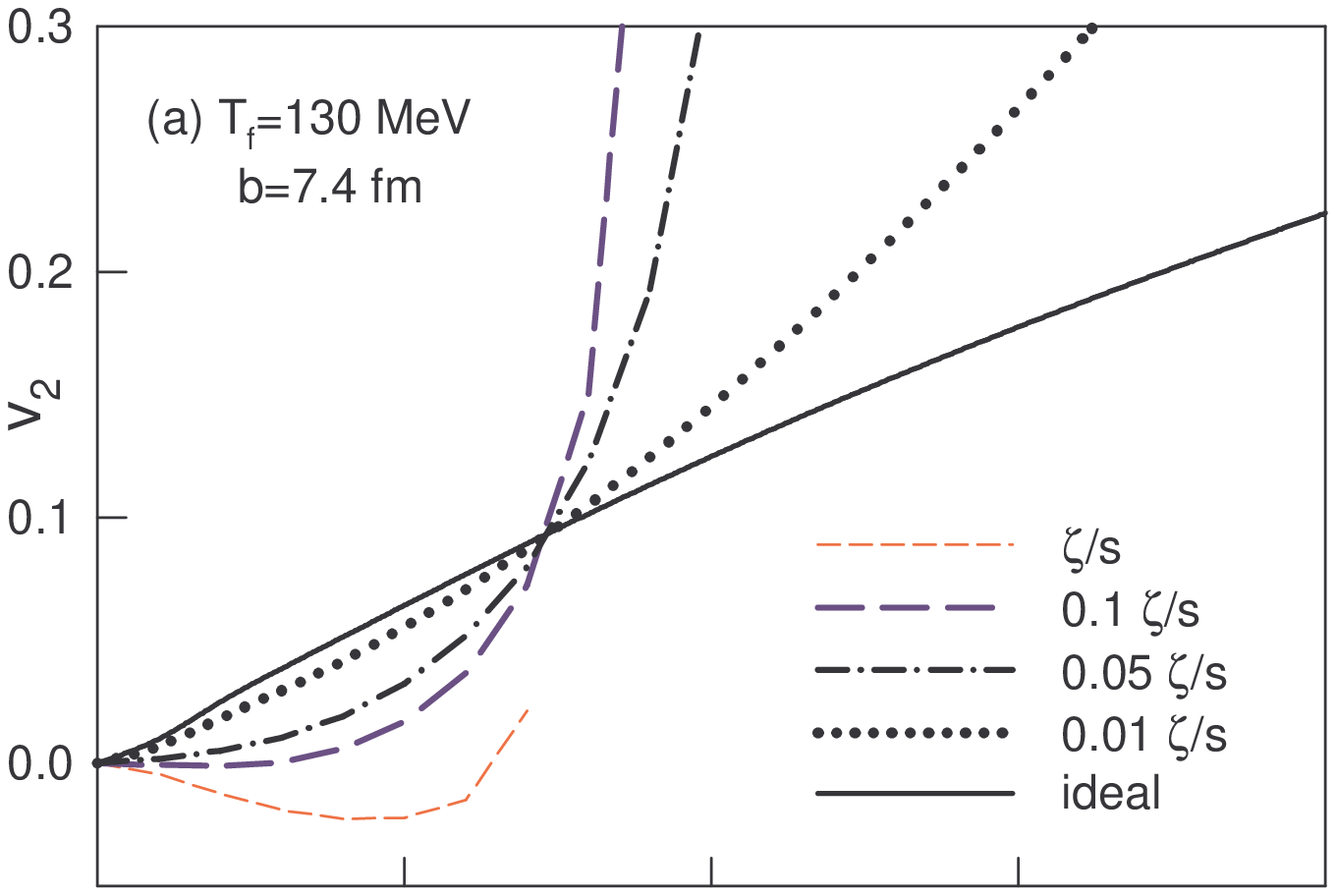}
		\includegraphics[scale=0.5]{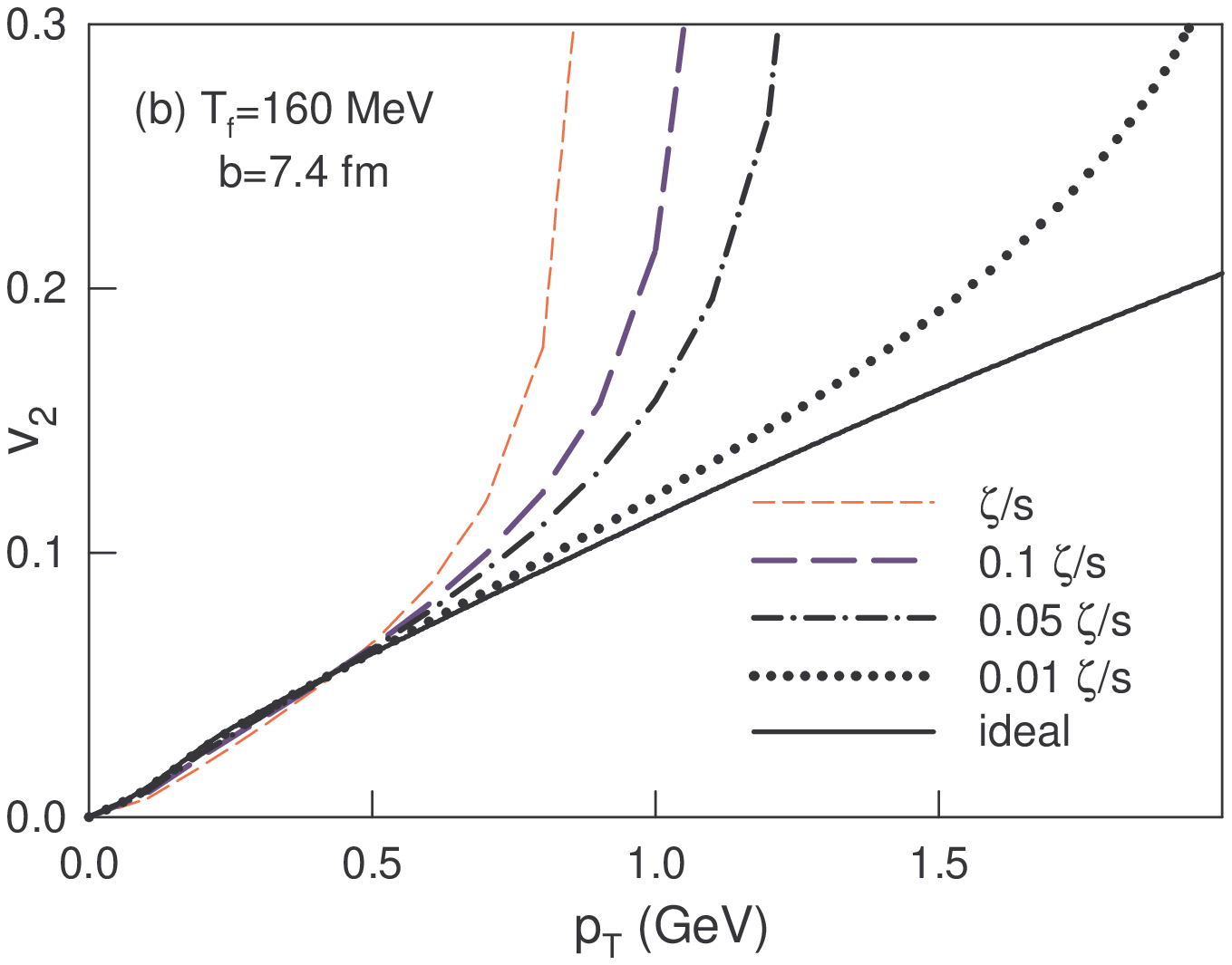}
	\caption{(Color online) Same as figure \ref{fig:grad_spectra} but for
	elliptic flow $v_{2}$.}
	\label{fig:deltaf_v2}
\end{figure}

\begin{figure}
	\centering
		\includegraphics[scale=0.5]{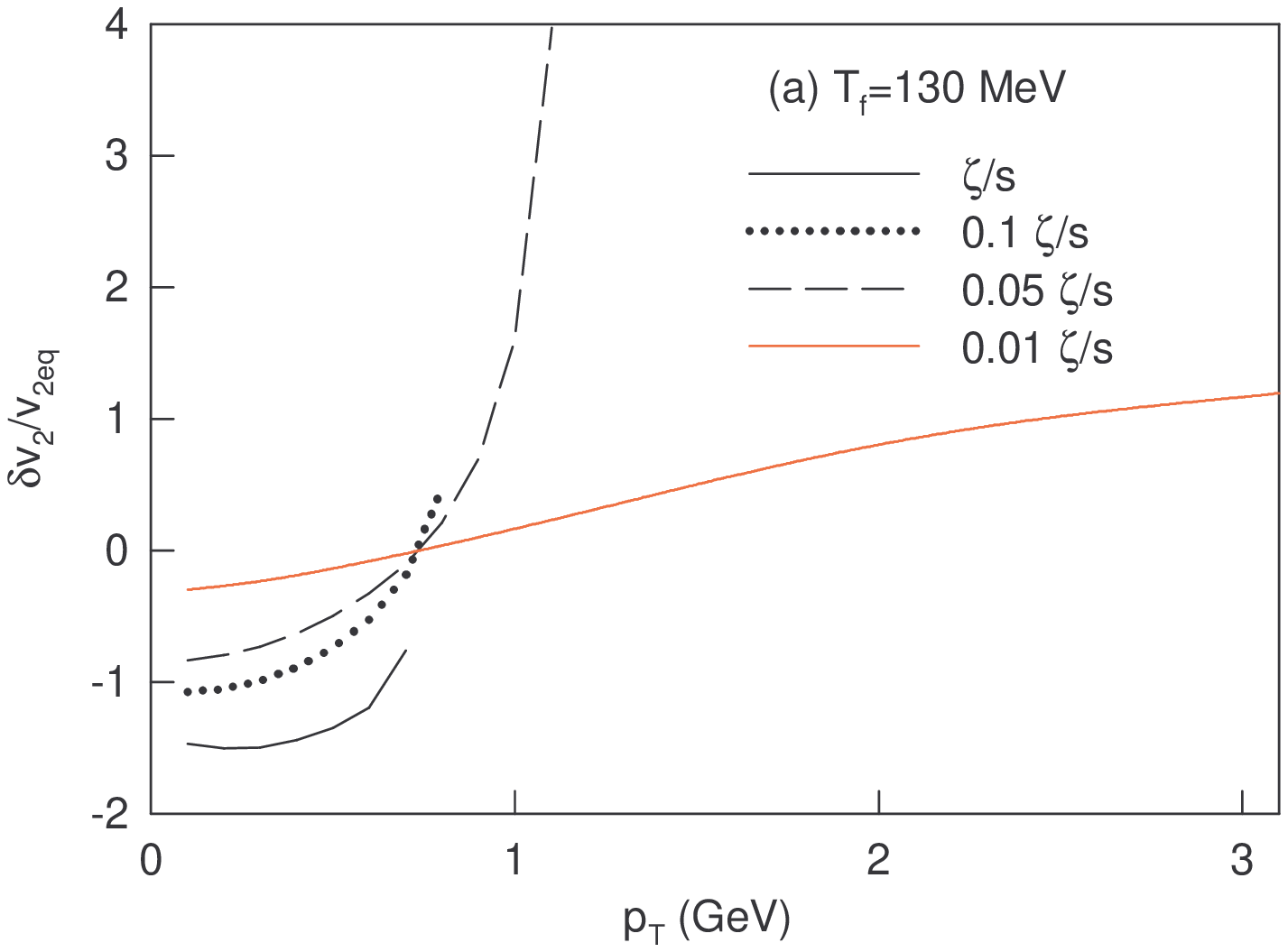}
		\includegraphics[scale=0.5]{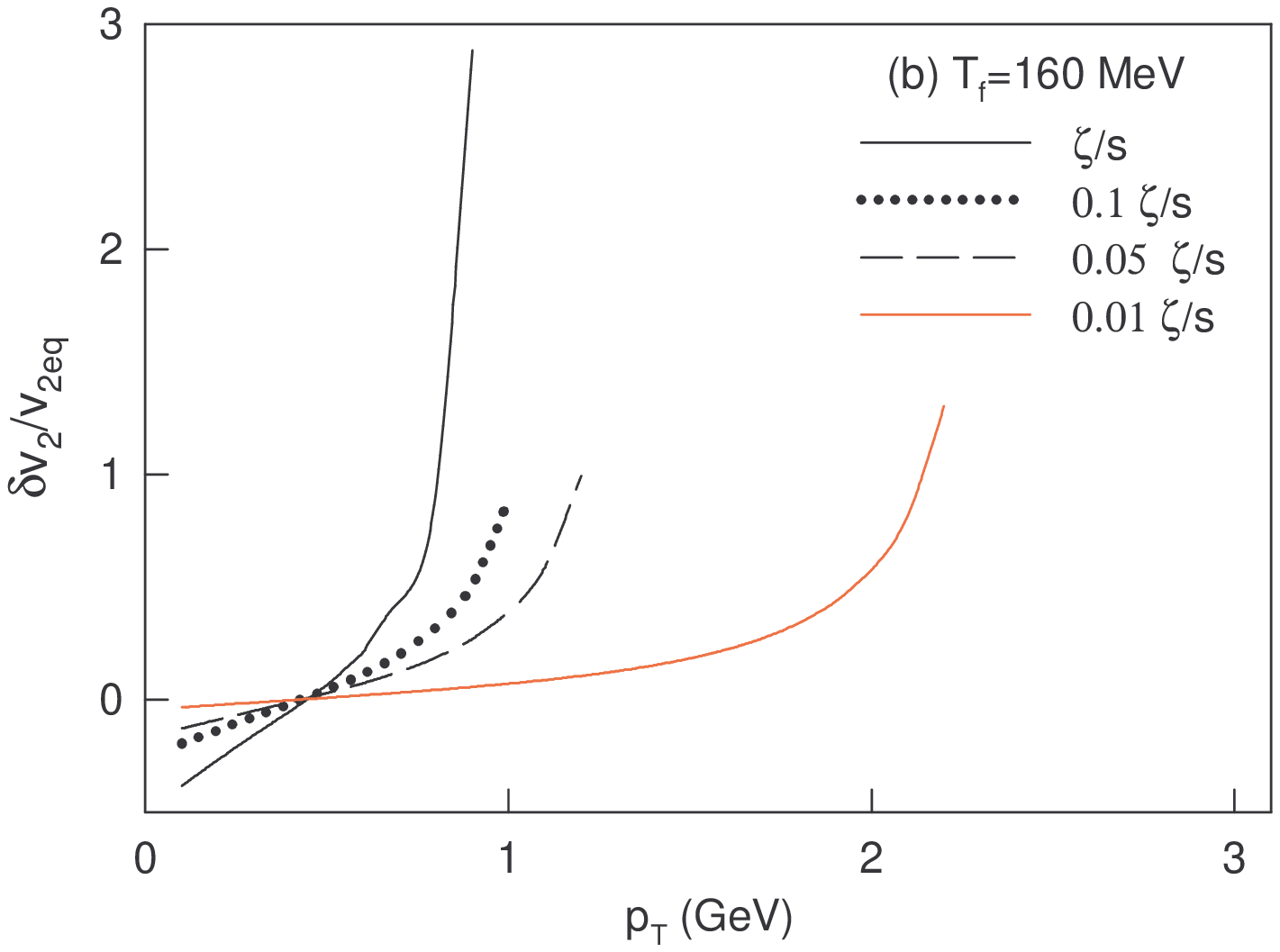}
	\caption{(Color online) Same as figure \ref{fig:grad_130} but for
	elliptic flow $v_{2}$.}
	\label{fig:correction_v2}
\end{figure}

Fig.~\ref{fig:deltaf_v2} shows the elliptic flow of pion for 20-30\% collision
centrality as a function of $p_{T}$ for ideal and bulk viscous evolution. The 
lines represents the same conditions as described in Fig.~\ref{fig:grad_spectra}.
The relative correction to the $v_{2}$ is defined in the same way as for the $p_{T}$
spectra is shown in Fig.~\ref{fig:correction_v2} (a) and (b) for $T_{f}$=130MeV and 
160 MeV respectively. The $v_{2}$ puts a more stricter constraint than $p_{T}$ spectra
on the allowed value of $\zeta/s$ for the applicability of Grad's 14 moment method.
We find the relative correction within 50\% for bulk viscosity to entropy density
ratio less than 0.05 times $\zeta/s$ form-1.  

\begin{figure}
	\centering
		\includegraphics[scale=0.5]{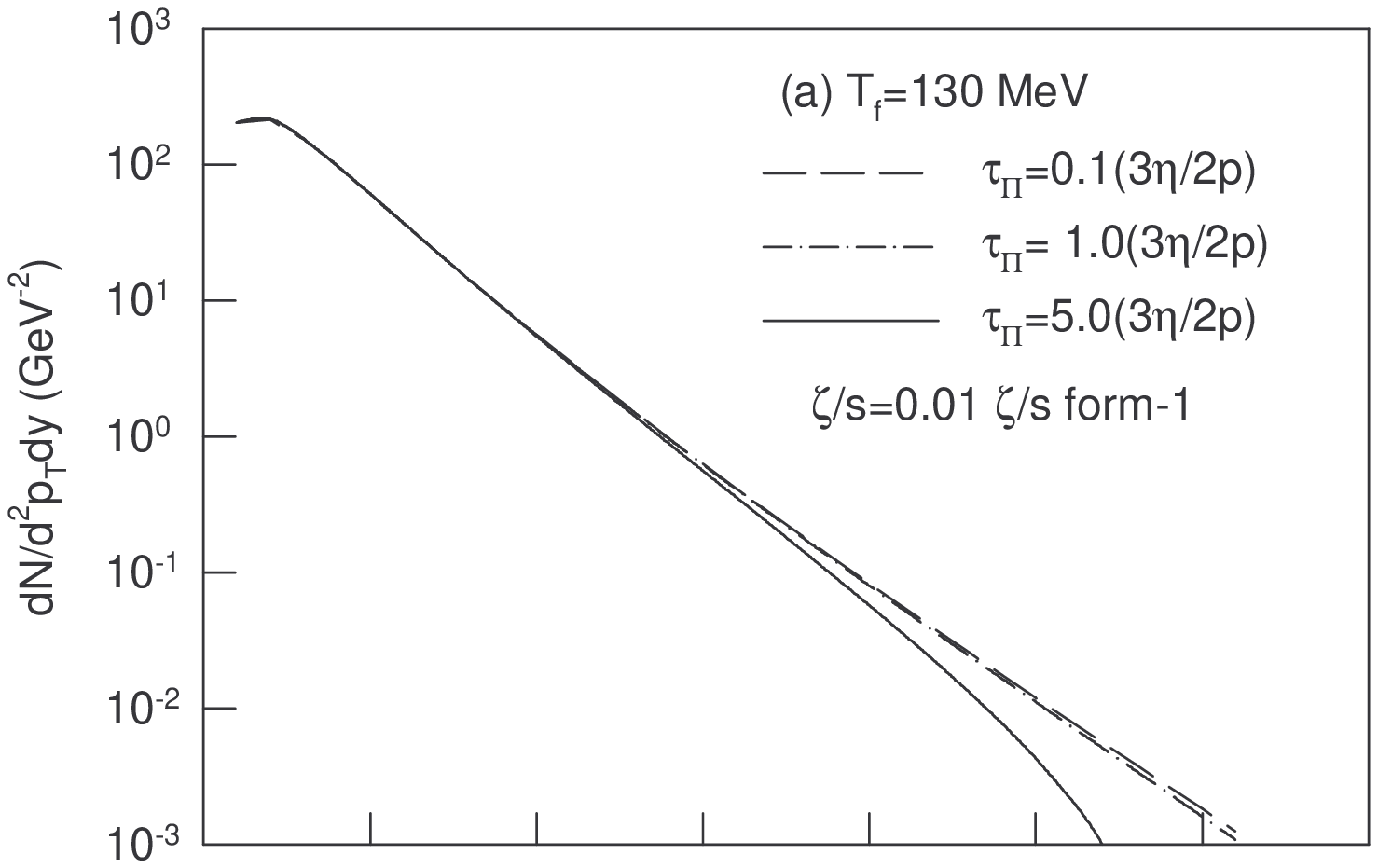}
		\includegraphics[scale=0.5]{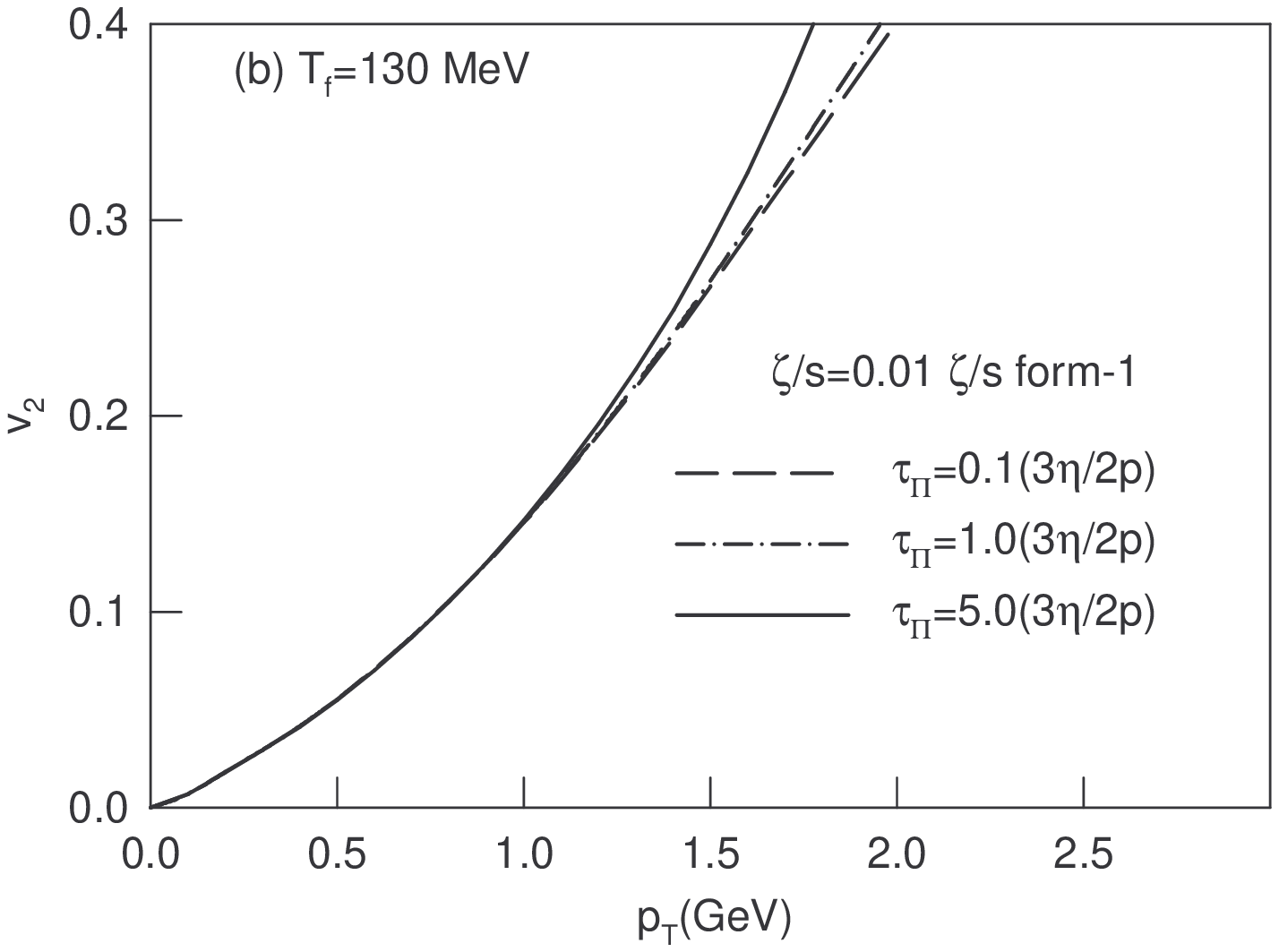}
	\caption{Effect of relaxation time on $p_T$ spectra and elliptic flow.}
	\label{fig:diff_relx}
\end{figure}

In all the above calculation, for the relaxation time of bulk viscosity, we have used the Boltzmann estimate for the relaxation time for shear viscosity  $\tau_\Pi=\tau_\pi$. We have investigated the effect of  relaxation time on the $p_T$ spectra and elliptic flow.  For 
relaxation time $\tau_\Pi$=0.1, 1 and 5 times $\tau_\pi$, we have solved the hydrodynamic evolution and computed $\pi^-$ $p_T$ spectra and elliptic flow. Results are shown in Fig.\ref{fig:diff_relx}.  If relaxation time is decreased by a factor for 10 from $\tau_\Pi=\tau_\pi$ to $\tau_\Pi=0.1\tau_\pi$, $p_T$ spectra or elliptic flow hardly changed. If relaxation time is increased by a factor of 5, at large $p_T$ yield is reduced marginally. Effect is more pronounced on elliptic flow. $v_2$ is increased. Increased flow with increasing relaxation time can be understood as follows: bulk stress evolve comparatively slowly with increased $\tau_\Pi$ and the non-equilibrium correction at the freeze-out is increased.
Increased non-equilibrium correction will lead to increased $v_2$.


\section{Summary}
\label{sec:Summary}

The effect of bulk viscosity on pion $p_{T}$ spectra and elliptic
flow was studied by numerically solving  2+1d  relativistic viscous hydrodynamics equations. 
Two different parametrize form of $\zeta/s$(T) was used along
with constant $\eta/s$. To construct $\zeta/s$ in the partonic phase we use lattice data.
$\zeta/s$ in hadronic phase is calculated using a hadron resonance gas model 
including hagedorn states with limiting mass of 80 GeV. 

A comparative study of ideal,shear and bulk viscous evolution is also done.
We observe that the time variation of the temperature of the fluid  remains 
similar for ideal and bulk viscous evolution.
Whereas in presence of shear viscosity the cooling rate is reduced.
Because of the reduced pressure due to finite $\zeta/s$ ,the transverse velocity 
slightly decreases compared to the ideal fluid around freezeout time. 
The shear viscosity on the other hand increases
the transverse pressure which results in a higher transverse velocity  compared to
the ideal evolution. The observable consequence of the above fact is reflected in the 
slope of the $p_{T}$ spectra of pion. The time evolution of spatial eccentricity
is almost unchanged in ideal and bulk viscous evolution. Due to the large transverse
velocity in shear viscous evolution the spatial deformation shows a rapid change compared
to ideal fluid. The momentum anisotropy in the shear viscous evolution grows at a much slower 
rate compared to ideal evolution. The change in the time variation of 
momentum anisotropy in the bulk viscous fluid with respect to ideal case is observed at 
a late time of the evolution. This will have consequences to the observed elliptic flow
of the hadrons.

The modification of the freezeout distribution
function according to Grad's moment method has been considered. 
The change in $p_{T}$ spectra and $v_{2}$ in bulk viscous evolution 
with respect to ideal simulation 
with no correction to the freezeout distribution is within 5-10\% 
depending on the form of $\zeta/s$(T).
However a large correction to both the $p_{T}$ spectra and elliptic flow
was observed for bulk viscous simulation with dissipative correction to the
freezeout distribution function. Combined study of $p_{T}$ spectra and $v_{2}$
of pion in 20-30\% Au-Au collision with full bulk viscous evolution puts
a constraints on the applicability of Grad's 14 moment method. 
We find the relative correction within 50\% for bulk viscosity to entropy density
ratio less than 0.01 times $\zeta/s$ form-1.

$\bf{Acknowledgment:}$
One of the author V.R would like to thank Rajeev Bhalerao,
Sourendu Gupta,Matthew Luzum,Bedangadas 
Mohanty and Jean Yves Ollitrault for their valuable comments,suggestion 
and helpful discussion during the work.

\appendix
\section{Evolution equations}
\label{Appendix1}
The energy momentum tensor including shear and bulk viscosity has the
following component in 2+1 dimension in $\tau,x,y,\eta$ co-ordinate.
\begin{eqnarray}
\nonumber
	T^{\tau\tau}=(\epsilon+p+\Pi)\gamma_{\bot}^{2}-(p+\Pi)+\pi^{\tau\tau}\\ \nonumber
	T^{x\tau}=(\epsilon+p+\Pi)\gamma_{\bot}^{2}v_{x}+\pi^{x\tau}\\ 
	T^{y\tau}=(\epsilon+p+\Pi)\gamma_{\bot}^{2}v_{y}+\pi^{y\tau}.
	\label{eq:energy_mom_tensor}
\end{eqnarray}
The conservation equations $\partial_{\mu}T^{\mu\nu}=0$ in 2+1 
dimension are written in the following forms to solve them numerically.
The three energy-momentum conservation equation along with four relaxation
equation(1 eq. for bulk and 3 eq. for shear viscosity) for viscous stress tensor
are solved by using the "`SHASTA"'\cite{Boris:shasta} algorithm. 
\begin{eqnarray}
\nonumber
	\partial_{\tau}\widetilde{T}^{\tau\tau}+\partial_{x}(\widetilde{T}^{\tau\tau}\overline{v}_{x})+\\
	\partial_{y}(\widetilde{T}^{\tau\tau}\overline{v}_{y})=-(p+\Pi+\tau^{2}\pi^{\eta\eta})
\end{eqnarray}

\begin{eqnarray}
\nonumber
\noindent	\partial_{\tau}\widetilde{T}^{\tau x}+\partial_{x}(\widetilde{T}^{\tau x}v_{x})+
	\partial_{y}(\widetilde{T}^{\tau x}v_{y})= \\
	-\partial_{x}(\widetilde{p}+\widetilde{\Pi}+\widetilde{\pi}^{xx}-v_{x}\widetilde{\pi}^{\tau x})-
	\partial_{y}(\widetilde{\pi}^{yx}-\widetilde{\pi}^{\tau x}v_{y})
\end{eqnarray}

\begin{eqnarray}
\nonumber
\noindent	\partial_{\tau}\widetilde{T}^{\tau y}+\partial_{x}(\widetilde{T}^{\tau y}v_{x})+
	\partial_{y}(\widetilde{T}^{\tau y}v_{y})= \\
	-\partial_{y}(\widetilde{p}+\widetilde{\Pi}+\widetilde{\pi}^{yy}-v_{y}\widetilde{\pi}^{\tau y})-
	\partial_{x}(\widetilde{\pi}^{xy}-\widetilde{\pi}^{\tau y}v_{x})
\end{eqnarray}

Where $\widetilde{T}^{\mu\nu}= \tau T^{\mu\nu}$ and $\widetilde{p}=\tau p$ , $\widetilde{\Pi}=\tau\Pi$,
$\overline{v}_{x}=\frac{T^{x\tau}}{T^{\tau\tau}}$, $\overline{v}_{y}=\frac{T^{y\tau}}{T^{\tau\tau}}$.

The relaxation equation for bulk and shear tensor takes the following form
\begin{eqnarray}
\nonumber
	\frac{\partial\Pi}{\partial\tau}+v_{x}\frac{\partial\Pi}{\partial x}+v_{y}\frac{\partial\Pi}
	{\partial y}=\\
	-\frac{1}{\tau_{\Pi}\gamma_{\tau}}[\Pi+\zeta\theta+\frac{1}{2} \Pi\tau_{\Pi}\partial_{\mu}u^{\mu}+
\zeta T \Pi D\left(\frac{\tau_{\Pi}}{\zeta T}\right)].
	\label{eq:rel_bulk_2}
\end{eqnarray}
\begin{eqnarray}
\label{eqc4a} \nonumber
\partial_\tau \pi^{xx} +v_x \partial_x \pi^{xx}+v_y \partial_y \pi^{xx}
&=&
-\frac{1}{\tau_\pi \gamma} \left (\pi^{xx} - 2\eta \sigma^{xx}\right )-\frac{1}{\gamma}I^{xx}_{1}\\
\label{eqc5} \nonumber
\partial_\tau \pi^{yy} +v_x \partial_x \pi^{yy}+v_y \partial_y \pi^{yy}
&=&
-\frac{1}{\tau_\pi \gamma} \left (\pi^{yy} - 2\eta \sigma^{yy}\right )-\frac{1}{\gamma}I^{yy}_{1}\\
\label{eqc6} \nonumber
\partial_\tau \pi^{xy} +v_x \partial_x \pi^{xy}+v_y \partial_y \pi^{xy}
&=&
-\frac{1}{\tau_\pi \gamma} \left (\pi^{xy} - 2\eta \sigma^{xy}\right )-\frac{1}{\gamma}I^{xy}_{1}
\end{eqnarray}
where 
 \begin{eqnarray}
\label{eqc7} \nonumber
\sigma^{xx}
=&&-\partial_x u^x -u^x Du^x -\frac{1}{3} \Delta^{xx} \theta\\ 
\label{eqc8} \nonumber
\sigma^{yy}
=&&-\partial_y u^y -u^y Du^y -\frac{1}{3} \Delta^{yy} \theta\\ 
\label{eqc9} \nonumber
 \sigma^{xy}
=&&-\frac{1}{2}[\partial_x u^y - \partial_y u^x
-u^x Du^y - u^y Du^x] \nonumber \\
&& -\frac{1}{3} \Delta^{xy} \theta
\end{eqnarray}
and

\begin{eqnarray} 
\nonumber
I^{xx}_{1}=
&&2u^{x}\left[\pi^{0x}Du_{0}-\pi^{xx}Du_{x}-\pi^{yx}Du_{y}\right] \nonumber \\
I^{yy}_{1}=
&&2u^{y}\left[\pi^{0y}Du_{0}-\pi^{xy}Du_{x}-\pi^{yy}Du_{y}\right] 
\end{eqnarray}
\begin{widetext}
\begin{eqnarray}
I^{xy}_{1}=\left(u^{x}\pi^{0y}+u^{y}\pi^{0x}\right)Du_{0}-\left(u^{x}\pi^{xy}+u^{y}\pi^{xx}\right)Du_{x}
-\left(u^{x}\pi^{yy}+u^{y}\pi^{yx}\right)Du_{y}
\end{eqnarray}
\end{widetext}

Here $D=u^\mu \partial_\mu$ is the convective time derivative and $\theta=\partial_\mu u^\mu$ is the expansion scalar. $\tau_\pi$ is the relaxation time for shear stress, $\tau_\pi=2\eta \beta_2$ and
$\tau_\Pi=\zeta\beta_{0}$ is the relaxation time for bulk viscous stress.

We solve only above three component of shear stress.
The dependent shear stress tensor components can be obtained from
the independent ones. Using the properties that (i) $\pi^{\mu\nu}$ is transverse to $u^\mu$ and (ii) $\pi^{\mu\nu}$ is traceless, $g^{\mu\nu}\pi_{\mu\nu}=0$,the dependent shear stress tensor components can be obtained as,
  
\begin{eqnarray}
\pi^{\tau x}=&& v_x \pi^{xx} + v_y \pi^{xy} \label{eqc1} \\
\pi^{\tau y}=&& v_x \pi^{xy} + v_y \pi^{yy} \label{eqc2}\\
\pi^{\tau \tau}=&& v^2_x \pi^{xx} + v^2_y \pi^{yy}+2v_x v_y \pi^{xy} \label{eqc3}\\
\tau^2 \pi^{\eta\eta}=&&   
-(1-v^2_x)\pi^{xx} - (1-v^2_y)\pi^{yy} \nonumber\\
&&+2v_xv_y\pi^{xy} \label{eqc4}
\end{eqnarray}
\noindent 
\section{Bulk viscous correction to freezeout distribution}
\label{sec:Appendix2}

In Cooper Frey prescription, the particle
momentum distribution is obtained by integrating the single particle distribution function over the freezeout hyper surface $\Sigma_{\mu}$.
\begin{equation}
	E\frac{dN}{d^{3}p}=\frac{g}{(2\pi)^{3}}\int d\Sigma_{\mu}p^{\mu}f(p^{\mu}u_{\mu},T),
\end{equation}

\noindent where E is the energy, g is degeneracy, and $p^{\mu}$ is four momentum.
In ideal hydrodynamics the fluid is in local thermal equilibrium and the distribution function
is  the one particle equilibrium distribution function $f_{eq}(x,p)$,

\begin{equation}\label{eq6_2}
f(x,p)=f^{eq}(x,p)=\frac{g}{2\pi^3}\frac{1}{exp[\beta(u_\mu p^\mu -\mu)] \pm 1}.
\end{equation} 

\noindent with inverse temperature $\beta=1/T$ and chemical potential $\mu$. $g$ is the degeneracy factor. The $(\pm)$ are respectively for fermions and bosons.
For dissipative fluids, the system is not in local thermal equilibrium. 
In a highly non-equilibrium system,   distribution
function $f(x,p)$ is unknown. 
If the system is slightly off-equilibrium,  then
it is possible to calculate correction to equilibrium distribution 
function due to   (small) non-equilibrium effects. Slightly
off-equilibrium distribution function can be  approximated  as,
 \begin{eqnarray}
f(x,p)=f_{eq}(x,p)+\delta f
\end{eqnarray}

Where $\delta f =\delta f_{bulk} + \delta f_{shear} << f$ represents the dissipative correction to the equilibrium 
distribution function $f_{eq}$, due to bulk viscosity and shear viscosity.

There are different methods available to calculate the dissipative correction
to the distribution function \cite{arXiv:1110.6742,Denicol:2009am,Dusling:2007gi,
Monnai:2009ad,arXiv:1109.5181,arXiv:1004.2023}. In order that the energy-momentum tensor remains continuous
across the freezeout surface the functional form of the $\delta f$ must be such that
the Landau matching condition should be satisfied; $u_{\mu}\Delta T^{\mu\nu}u_{\nu}=0$
\cite{arXiv:1004.2023}. For bulk viscosity,
the dissipative correction for a multicomponent system was calculated by using Grad's
fourteen moment method in \cite{Monnai:2009ad}. Following \cite{Monnai:2009ad} the
dissipative correction for bulk viscosity $\delta f_{bulk}$ can be written in the
following way,  

 \begin{eqnarray}
\nonumber
\delta f_{bulk}=-f_{eq}(1+\epsilon f_{eq}) \times \\ \nonumber
\left[D_{0}p^{\mu}u_{\mu}+B_{0}p^{\mu}p^{\nu}\Delta_{\mu\nu}+\tilde{B}_{0}p^{\mu}p^{\nu}u_{\mu}u_{\nu}\right]\Pi	
\end{eqnarray}

Where the prefactor $D_{0},B_{0}$ and $\tilde{B}_{0}$ are temperature dependent constant. 
Here we have dropped the index 'i' of particle species for simplicity.
The Landau matching condition is satisfied with the present form of correction to the 
ideal distribution function.
In \cite{Monnai:2009ad} the prefactors $D_{0},B_{0}$ and $\tilde{B}_{0}$ was calculated for a multicomponent 
hadron gas. We use their estimated values(given in the table~\ref{table2}) in this calculation for two different freezeout 
temperature $T_{f}=130 MeV$ and $T_{f}=160 MeV$. 
\begin{table}[h]
\caption{\label{table2} Prefactors for two different temperature} 
\begin{ruledtabular} 
  \begin{tabular}{|c|c|c|c|}
 $T_{f}$         & $D_{0}(GeV^{-5})$ & $B_{0}(GeV^{-6})$& $\tilde{B_{0}}(GeV^{-6})$   \\ \hline
  130 MeV        & 9.10$\times10^{4}$ &1.12$\times10^{5}$ &-3.27$\times10^{4}$ \\ \hline
  160 MeV        & 2.01$\times10^{4}$  & 1.66$\times10^{4}$ &-7.84$\times10^{3}$   \\ 
 \end{tabular}\end{ruledtabular}  
\end{table} 

Shear viscous correction to equilibrium distribution function is well known,

\begin{equation} \label{eqv5}
\delta f_{shear} =f_{eq}(1+\epsilon f_{eq})\frac{1}{2(\varepsilon+p)T^2} p_\mu p_\nu \pi^{\mu\nu},
\end{equation}

The correction to the ideal spectra $\frac{dN(ideal)}{d^{2}p_{T}dy}$ due to the bulk viscosity is

\begin{equation}
\frac{d\delta N(bulk)}{d^{2}p_{T}dy}=\frac{g}{\left(2\pi\right)^{3}}\int_{\Sigma}d\Sigma_{\mu}p^{\mu}\delta f_{bulk}(p^{\mu}u_{\mu},T)
\label{eq:deltaf}
\end{equation}
Now the four momentum of the fluid element is $p^{\mu}=(m_{T}coshy,p_{x},p_{y},m_{T}sinhy)$ where
$m_{T}=\sqrt{m^{2}_{0}+p^{2}_{T}}$ and the momentum rapidity is $y=\frac{1}{2}ln\frac{E+p_{z}}{E-p_{z}}$.
The freeze-out hypersurface in $(\tau,x,y,\eta)$ co-ordinate is
\begin{eqnarray}
\nonumber
d\Sigma_{\mu}=\left(m_{T}cosh\eta, -\frac{\partial\tau_{f}}{\partial x},-\frac{\partial\tau_{f}}{\partial y}m_{T}sinh\eta\right)\tau_{f}dxdyd\eta
\end{eqnarray}
and
\begin{eqnarray}
\nonumber
p^{\mu}.d\Sigma_{\mu}=\left(m_{T}cosh(\eta-y)-\vec{p}_{T}.\vec{\nabla}_{T}\tau_{f}\right)\tau_{f}dxdyd\eta
\end{eqnarray}

Using these relationship into equation \ref{eq:deltaf} and after some algebra
we have the final form of the correction to the invariant yield due to the bulk viscosity
which is given as

\begin{widetext}
\begin{eqnarray}
\nonumber
\frac{d\delta N(bulk)}{d^{2}p_{T}dy}=A[m_{T}\left\{\frac{b_{1}}{4}k_{3}(n\beta_{\bot})+
\frac{3b_{1}}{4}k_{1}(n\beta_{\bot})
+\frac{b_{2}}{2}k_{0}(n\beta_{\bot})+ 
\frac{b_{2}}{2}k_{0}(n\beta_{\bot})+b_{3}k_{1}(n\beta_{\bot})\right\}\\ \nonumber
-\vec{p}_{T}.\vec{\nabla}_{T}\tau_{f}\left\{\frac{b_{1}}{2}k_{0}(n\beta_{\bot})
+\frac{b_{1}}{2}k_{2}(n\beta_{\bot})+b_{2}k_{1}(n\beta_{\bot})
+b_{3}k_{0}(n\beta_{\bot})\right\} ]\Pi	
\end{eqnarray}
\end{widetext}

where
$\beta_{\bot}=m_{T}\gamma\beta$

\begin{widetext}
\begin{eqnarray}
\nonumber
\noindent A&=&2\sum^{1}_{\infty}(\mp1)^{n+1}e^{n\beta\left(\gamma \vec{v}_{T}\vec{p}_{T}\right)}\\
 \nonumber
b_{1}&=&m^{2}_{T}\left(\tilde{B}_{0}-B_{0}\right)\gamma^{2}\\
 \nonumber
b_{2}&=&m_{T}\left\{D_{0}\gamma+2\gamma^{2}\left(p^{x}v_{x}+p^{y}v_{y}\right)B_{0}-
2\tilde{B}_{0}\gamma^{2}\left(p^{x}v_{x}+p^{y}v_{y}\right)\right\}\\
\nonumber
b_{3}&=&B_{0}m^{2}_{T}-D_{0}\gamma\left(p^{x}v_{x}+p^{y}v_{y}\right)-B_{0}p^{2}_{x}
\left(1+\gamma^{2}v^{2}_{x}\right)-B_{0}p^{2}_{y}\left(1+\gamma^{2}v^{2}_{y}\right)\\ \nonumber
&-&2B_{0}p^{x}v_{x}p^{y}v_{y}\gamma^{2}+\tilde{B}_{0}\gamma^{2}\left(p^{x}v_{x}\right)^{2}
+2\tilde{B}_{0}\gamma^{2}p^{x}v_{x}p^{y}v_{y}+\tilde{B}_{0}\gamma^{2}\left(p^{y}v_{y}\right)^{2}
\end{eqnarray}
\end{widetext}


\end{document}